\documentclass[%
 reprint,
 amsmath,amssymb,
 aps,
]{revtex4-2}

\usepackage{graphicx}
\usepackage{dcolumn}
\usepackage{bm}

\begin{document}

\preprint{APS/123-QED}

\title{Probing black bounce spacetime through mass accretion rate}

\author{Xi-Rui Guo}
\email{xiruiguo@nuaa.edu.cn}
 \affiliation{Center for the Cross-disciplinary Research of Space Science and Quantum-technologies (CROSS-Q), College of Physics, Nanjing University of Aeronautics and Astronautics, Nanjing, 210016, China}
\author{Ya-Peng Hu}%
 \email{huyp@nuaa.edu.cn}
\author{Yu-Sen An}
\email{anyusen@nuaa.edu.cn}
\thanks{Correspondence author}
\affiliation{Center for the Cross-disciplinary Research of Space Science and Quantum-technologies (CROSS-Q), College of Physics, Nanjing University of Aeronautics and Astronautics, Nanjing, 210016, China
}
\affiliation{
 MIIT Key Laboratory of Aerospace Information Materials and Physics,  Nanjing University of Aeronautics and Astronautics, Nanjing, 210016, China
}%

\date{\today}

\begin{abstract}

We study steady, spherically symmetric accretion of a collisionless kinetic gas onto a compact object described by the Simpson--Visser (SV) metric, a one-parameter regularization of the Schwarzschild spacetime. Using the standard relativistic Vlasov 
formalism, we adopt boundary conditions at a finite injection radius $R$ and compute the mass accretion rate $\dot{M}$ for Vlasov gas with both monoenergetic and Maxwell--J\"uttner distribution functions. Throughout the computation, we also take into account the influence of particle's angular momentum. 
We carefully investigate the influence of the black bounce parameter on the mass accretion rate $\dot{M}$. By expanding accretion rate in terms of bounce parameter $a$, we found that the leading-order correction appears at quadratic order in $a$. This work extends the Vlasov accretion framework to a broader class of spherically symmetric spacetimes and quantifies the impact of the geometric modification on the accretion flow. In particular, since the shadows of the SV and Schwarzschild black holes are degenerate, comparing their mass accretion rates may offer a complementary probe for distinguishing these geometries in future astrophysical observations.
\end{abstract}

\maketitle


\section{Introduction}
\label{sec:Introduction}

Accretion of matter onto compact objects is a fundamental astrophysical process that governs the energy output of active galactic nuclei \cite{1998agn..book.....K}, X-ray binaries \cite{1995xrbi.nasa.....L}, and the growth of supermassive black holes \cite{Kormendy:2013dxa}. The standard theoretical framework for spherical accretion was developed for hydrodynamic fluids by Bondi \cite{Bondi:1952ni} and Hoyle and Lyttleton \cite{Hoyle_Lyttleton_1939}, and later generalized to general relativity by Michel \cite{Michel:1972oeq}. These fluid models assume that the accreting gas is sufficiently collisional to maintain local thermal equilibrium, which is a good approximation for dense, cool flows. However, in low-luminosity systems such as Sgr A* and M87*, the plasma is hot and dilute, with a mean free path comparable to or larger than the gravitational scale (see, e.g., Ref.~\cite{Yuan:2014gma,Mahadevan_1997,Baganoff_2003,EventHorizonTelescope:2019dse}). In such environments, collisions are rare and the gas behaves effectively as a collisionless kinetic system (see, e.g., Ref.~\cite{2002rbet.book.....C,Rioseco:2016jwc,Gamboa:2021ytn}), requiring a description based on the Vlasov  equation (collisionless Boltzmann equation, e.g., Ref.~\cite{2002rbet.book.....C}) rather than hydrodynamics simulations.

Gamboa et al.~\cite{Gamboa:2021ytn} recently studied steady spherical accretion of a Vlasov gas onto a Schwarzschild background. A key innovation of their work was to set boundary conditions at a finite radius $R_s$ (the "injection sphere" in Schwarzschild background) rather than at infinity, as a real astrophysical observations provide estimates of density and temperature at a finite distance from the black hole. They showed that when individual particles are allowed to have angular momentum, the resulting mass accretion rate is significantly suppressed compared to the purely radial case and can be several orders of magnitude lower than the Bondi rate. Their formalism provides a systematic way to compute $\dot{M}$ for any static, spherically symmetric metric, once the geodesic structure is known.

The formalism developed in Ref.~\cite{Gamboa:2021ytn} is not restricted to the Schwarzschild black hole and can be applied to any static, spherically symmetric black hole. Among numerous black hole solutions, regular black holes, in which the central singularity is removed by additional matter fields or modified gravity effects, have attracted considerable attentions. This class of black holes was first proposed by Bardeen \cite{1968qtr..conf...87B} and later developed by many researches such as \cite{Hayward:2005gi,Roman:1983,Frolov:2014jva,Frolov:2016pav,Frolov:2017rjz}; they have grown up as an important part of black hole physics. Furthermore, the work of Morris et al.~\cite{10.1119/1.15620,PhysRevD.39.3182,PhysRevLett.61.1446} introduced another class of spacetimes in which the singularity is removed, the event horizon disappears, and a throat connecting distant regions forms at the center. Although these configurations remain theoretical, they have attracted substantial interest. Among these exotic solutions, the Simpson–Visser (SV) black hole solution \cite{Simpson:2018tsi,Simpson:2019cer,PhysRevD.103.084052} is particularly notable. It can be regarded as a one-parameter deformation of Schwarzschild, controlled by the bounce parameter $a$, a real parameter with dimensions of length (see Sec. \ref{sec:Black bounce spacetime} for details). The concept of a bounce parameter has precursors in the quantum-gravity literature \cite{Kazakov:1993ha}, in which Kazakov and Solodukhin argued that Planck-scale metric fluctuations modify the classical gravitational field equations and can yield a “quantum-deformed” Schwarzschild solution. In contrast, Simpson and Visser adopted a more phenomenological approach, introducing the bounce parameter $a$ to construct a one-parameter family of geometries that smoothly interpolates between a regular black hole and a traversable wormhole. These features make the SV spacetime a valuable framework for studying regular black holes and, more broadly, wormhole-like geometries. However, it remains difficult for current observations to distinguish the SV black hole from other black hole solutions, which poses a challenge for further investigation \cite{Shaikh:2022ivr,Islam:2021ful}. The well known black shadow observation fails to distinguish the SV spacetime from Schwarzschild case due to the degeneracies in the computation of impact factor. However, since particle geodesics differ among different black hole spacetimes, the corresponding accretion rates should differ from each other which may provide an alternative probe which can break this degeneracy.  Motivated by this, we calculate accretion rates in different spacetimes to identify distinctive SV features and to offer a new way of distinguishing between black hole solutions. In particular, applying the formalism of Ref.~\cite{Gamboa:2021ytn} to the SV black hole shows how the accretion rate depends on the bounce parameter, providing a useful probe of accretion onto different black holes and a potential means of discriminating among them.

In this work, we adopt the same assumptions as in Ref.~\cite{Gamboa:2021ytn} and discuss particle motion in the SV spacetime both with and without angular momentum. Using the monoenergetic and the Maxwell–Jüttner distribution functions, we calculate the corresponding mass accretion rate formulas for a kinetic gas. Unlike the results in Ref.~\cite{Gamboa:2021ytn}, the black hole mass accretion rate obtained here depends on the bounce parameter $a$. According to the work of Riaz et al. and others \cite{Riaz:2022rlx,Tripathi_2021,Bambi_2022}, X-ray observations constrain the bounce parameter to a small range $0\le a\le 0.49M$. This motivates us to expand the mass accretion rate expression around $a=0$ to show the difference between Schwarzschild black hole. Finally, since Sgr A* provides the most stringent observational constraints on the environment near a supermassive black hole, we adopt the same environmental parameters as in Ref. \cite{Shcherbakov:2010yq} but evaluate the mass accretion rate at the SV injection radius $R$, which is determined by fixing the proper radial distance from the horizon to ensure a consistent comparison with the Schwarzschild case. We investigate the extent to which the bounce parameter affects the mass accretion rate and assess the error of the expanded expression relative to the original formula. This allows us to examine whether the mass accretion rate can serve as a complementary probe for distinguishing SV black holes from Schwarzschild black holes.

This manuscript is organized as follows. In Section \ref{sec:Black bounce spacetime}, we introduce the SV spacetime geometry and the physical quantities of interest, including the horizon radius, the innermost stable circular orbit (ISCO), the effective potential of particles, and the determination of the SV injection radius $R$ by fixing the proper radial distance from the horizon. In Section \ref{sec:Vlasov Gas Accretion In The Simpson-Visser Background}, we first derive the mass accretion rate formula for the case of radial particle infall as an illustrative example, showing that the situation considered here is analogous to the Schwarzschild case. We then consider infalling particles with nonzero angular momentum and discuss the mass accretion rate formulas for particle distribution functions given by a mono-energetic distribution and a Maxwell–Jüttner distribution, respectively. Finally, we expand the two mass accretion rate expressions around $a=0$, substitute the environmental parameters, and numerically evaluate the mass accretion rate. In Section \ref{sec:Conclusion and Discussion}, we list and compare the numerical values of the mass accretion rate obtained in this work with those obtained for the Schwarzschild case in Ref.~\cite{Gamboa:2021ytn}. We discuss the feasibility of distinguishing Schwarzschild black holes from SV black holes by means of the mass accretion rate, and present a summary and outlook. Throughout this paper, we adopt geometrized units with $G=c=1$.

\section{Black bounce spacetime}
\label{sec:Black bounce spacetime}

In this section, we introduce the Simpson–-Visser (SV) black-bounce spacetime and discuss the properties relevant to the subsequent accretion analysis \cite{Simpson:2018tsi,Simpson:2019cer,PhysRevD.103.084052}. 

The line element is given by
\begin{equation}
ds^{2}=-\alpha(r)^{2}dt^{2}+\frac{dr^{2}}{\alpha(r)^{2}}+\rho^{2}\left(d\theta^{2}+\sin^{2}\theta d\phi^{2}\right),
\label{eq:line_element}
\end{equation}
where
\begin{align}
    \alpha(r)^2=1-\frac{2M}{\rho},\quad  \rho=\sqrt{r^2+a^2}
\end{align}
Here $M$ denotes the mass of the black hole, $a$ is the bounce parameter, a real parameter with dimensions of length due to the modified gravity effect. The metric is manifestly static and spherically symmetric. For $a=0$, the metric reduces to the Schwarzschild solution, whereas for $a\ne 0$, the geometry is regular everywhere and the parameter $a$ controls the nature of the spacetime. For $a<2M$, the horizon locations are determined by $g^{rr}=0$, which gives a pair of horizon
\begin{equation}
    r_\pm = \pm\sqrt{(2M)^2 - a^2}.
\end{equation}
Since the geometry is now everywhere regular, the spacetime does not terminate at the original singularity $r=0$ but instead can be extended to the region $r<0$. The hypersurface $r=0$ is a spacelike spherical surface which marks the boundary between our universe and a bounce into a separate copy of our own universe. For $a=2M$, the horizon shrinks to $r=0$ and becomes null and extremal, giving a one-way wormhole. For $a>2M$, no horizon exists, and the geometry describes a traversable wormhole whose throat is located at $r=0$.

We now turn to geodesic motions around the Simpson-Visser geometry. An overdot denotes differentiation with respect to an affine parameter $\lambda$. Because of spherical symmetry, we may restrict the motion to the equatorial plane $\theta = \pi/2$. The Lagrangian $\mathcal{L}$ for a test particle is
\begin{equation}
    \mathcal{L} = \frac12 g_{\mu\nu} \dot{x}^\mu \dot{x}^\nu = \frac12\left[ -\alpha(r)^2\dot{t}^2 + \frac{\dot{r}^2}{\alpha(r)^2} + (r^2+a^2)\dot{\phi}^2 \right].
\end{equation}
The conserved energy $E$ and angular momentum $L$ follow from the Killing symmetries,
\begin{align}
p_t &= \frac{\partial \mathcal{L}}{\partial \dot{t}} = -\alpha(r)^2\dot{t} = -E,\\
p_\phi &= \frac{\partial \mathcal{L}}{\partial \dot{\phi}} 
= (r^2 + a^2)\dot{\phi} = L.
\end{align}
Hence
\begin{align}
\dot{t} &= \frac{E}{\alpha(r)^2}, \\
\dot{\phi} &= \frac{L}{r^2 + a^2}.
\label{eq:conserved}
\end{align}
The normalization condition is
\begin{equation}
g_{\mu\nu}\dot{x}^\mu\dot{x}^\nu = \epsilon,
\label{eq:normalization}
\end{equation}
where $\epsilon=-1$ for massive particles and $\epsilon=0$ for massless particles.

Substituting Eq.~(\ref{eq:conserved}) into Eq.~(\ref{eq:normalization}), we obtain the radial equation
\begin{equation}
\dot{r}^2 = E^2 - V(r),
\end{equation}
where we define the effective potential $V(r)$ by
\begin{equation}
V(r) = \alpha(r)^2\left[-\epsilon + \frac{L^2}{r^2 + a^2}\right].
\end{equation}
The inner-most stable circular orbit(ISCO) satisfy $\frac{d}{dr}V|_{\epsilon=-1}=0$, which gives
\begin{equation}
r_{\rm ISCO} = \sqrt{(6M)^2 - a^2}.
\end{equation}
while the circular photon orbits satisfy $\frac{d}{dr}V|_{\epsilon=0}=0$, which gives
\begin{equation}
r_{\rm ph} = \sqrt{(3M)^2 - a^2}.
\end{equation}
Although the above quantities depend on $a$, determining whether the Schwarzschild and SV black holes can be distinguished through their shadows requires evaluating the shadow radius $r_{sh}$ \cite{EventHorizonTelescope:2022wkp,Vagnozzi:2022moj,Cunha:2018acu,Dokuchaev:2019jqq,Perlick:2021aok}. For a distant observer, the shadow radius of SV spacetime is given by 
\begin{equation}
r_{sh} = \sqrt{\frac{r_{\rm ph}^2 + a^2}{\alpha(r_{\rm ph})^2}}= 3\sqrt{3}M,
\end{equation}
This value is exactly the same as that for the Schwarzschild black hole. Consequently, the shadow radius is independent of $a$ 
thus shadow observations alone cannot distinguish the SV black hole from Schwarzschild black hole. This degeneracy motivates the use of the mass accretion rate as a complementary probe.

To compare the accretion flow in the two spacetimes on the same physical
footing, we should impose the injection sphere at the same proper radial distance from the horizon for both two cases. In the Schwarzschild spacetime computation, Ref.~\cite{Gamboa:2021ytn} choose to impose the injection sphere at the coordinate radius $R_s=5r_s=10M$, where $r_s=2M$ is the
Schwarzschild radius. However, as the SV geometry differs, the same 
coordinate radius does not correspond to the same physical proper distance to the horizon.
The proper radial distance from the horizon to the injection sphere 
in the Schwarzschild spacetime is
\begin{equation}
I_{Sch} = \int_{r_s}^{R_s} \frac{dr}{\sqrt{1-r_s/r}},
\label{eq:IR}
\end{equation}
while the corresponding quantity in the SV spacetime is
\begin{equation}
I_{SV} = \int_{r_H}^{R} \frac{dr}{\sqrt{1-r_s/\sqrt{r^2+a^2}}},
\label{eq:IL}
\end{equation}
where $r_H = \sqrt{r_s^2-a^2}$ is the SV horizon radius. For $a\ll r_s$, 
expanding $I_{SV}$ through second order in $a$ and imposing $I_{Sch}=I_{SV}$ gives
\begin{equation}
R \simeq R_{s} - \left(\frac{1}{r_s} - \frac{1}{2R_{s}}\right)a^2 + \mathcal{O}\left(a^4 \right).
\label{eq:R'}
\end{equation}
The derivation is presented in Appendix \ref{app:injection_sphere}.

\begin{figure}
    \centering
    \includegraphics[width=\columnwidth]{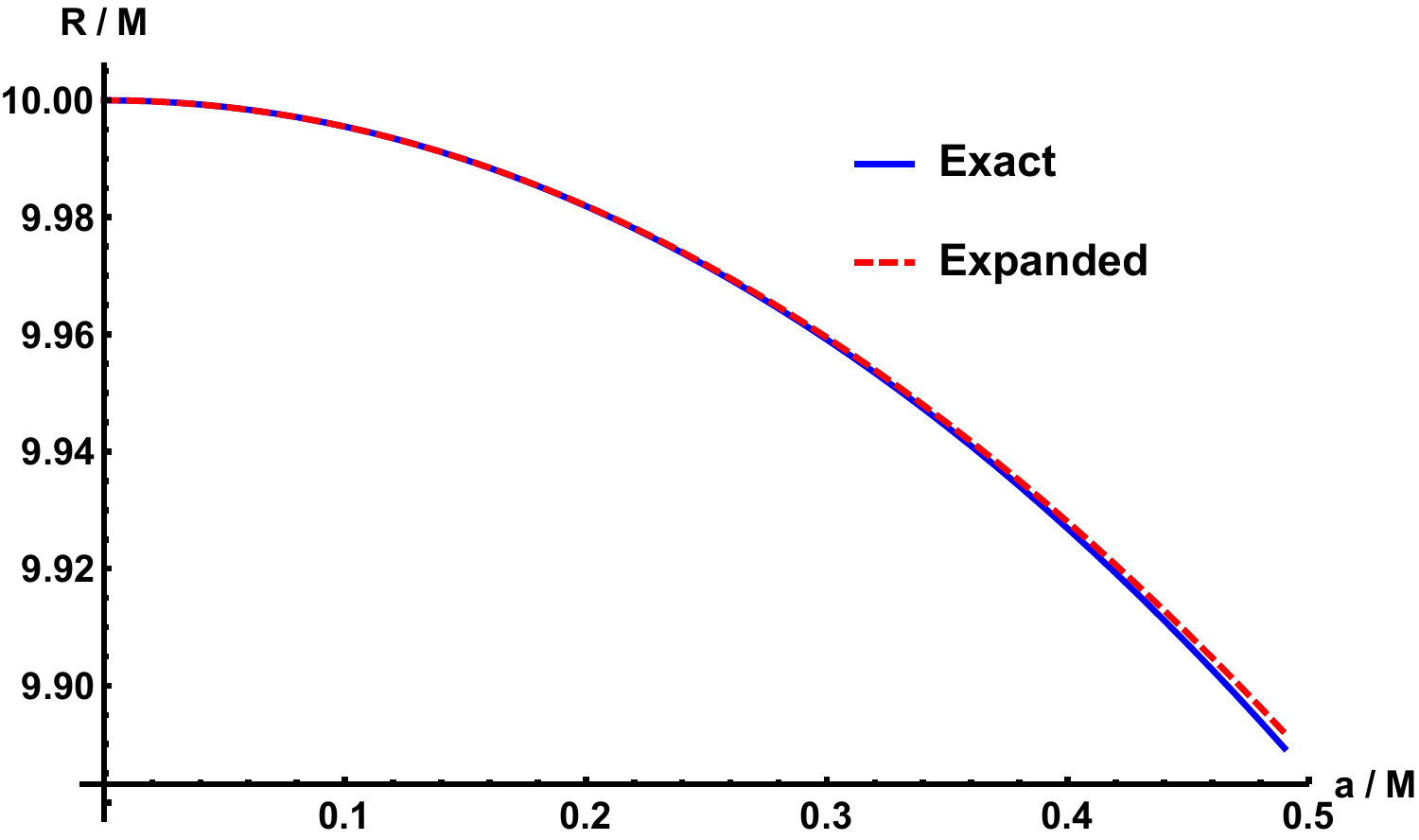}
    \caption{The injection radius $R$ as a function of $a/M$ for Simpson-Visser black hole case. The blue solid line denotes the exact numerical solution $R_{\rm exact}$, while the red dashed line represents the second-order analytical expansion $R^{(2)}$. The two curves are practically indistinguishable over the observationally allowed range $0 \le a/M \le 0.49$.}
    \label{fig:Ra}
\end{figure}

To assess the accuracy of the truncated expansion, we solve Eq.(\ref{eq:IL})
numerically with $I_{Sch}=I_{SV}$ and denote the solution by $R_{\rm exact}$.
Figure~\ref{fig:Ra} compares this exact solution with the second-order
result $R^{(2)}$ in Eq.~(\ref{eq:R'}); the two curves are practically
indistinguishable over the entire observationally allowed range
$0\leq a/M\leq 0.49$. The residual difference is quantified by the relative error
\begin{equation}
\Delta_1\equiv
\frac{|R_{\rm exact}-R^{(2)}|}{R_{\rm exact}}.
\label{eq:DeltaR}
\end{equation}
For $0\le a/M\le0.49$, $\Delta_1$ remains below $2.65\times10^{-4}$;
the corresponding curve is shown in Fig.~\ref{fig:Rerror} of
Appendix~\ref{app:injection_sphere}.

As a further check, including the $a^4$ term from Appendix~A reduces
$\Delta_1$ to approximately $10^{-5}$, while the $a^4$ contribution itself
accounts for only about $2.3\%$ of the leading $a^2$ shift. These results
indicate that the truncation error is well controlled over the range of
interest and support retaining terms through second order in the
subsequent derivation of the mass accretion rate.

For $R_s=10M$ and $a/M=0.49$, Eq.~(\ref{eq:R'}) gives
$R\simeq 4.945\,r_s$, which is only about $1.1\%$ smaller than $R_s$.
We therefore assume that the environmental parameters at the injection
sphere, such as the temperature, particle velocity, and number density,
can be taken as those specified at $R_s$ in Ref.~\cite{Gamboa:2021ytn}. 
In the numerical evaluation we use $R_{\rm exact}$, 
while the analytical expressions below
are written in terms of a generic radius $R$. The remaining environmental
parameters are kept as given by Baganoff et al.~\cite{Shcherbakov:2010yq,Baganoff_2003}.

According to this comparison method, we will investigate the mass accretion phenomenon in SV background in the next section. 

\section{Vlasov Gas Accretion In The Simpson-Visser Background}
\label{sec:Vlasov Gas Accretion In The Simpson-Visser Background}
For the strong gravitational field near a black hole, the accretion flow 
approaches relativistic speeds, necessitating a fully relativistic kinetic 
description. In curved spacetime, the phase space naturally acquires the 
structure of a cotangent bundle $T^*\mathcal{M}$, with local coordinates 
$(x^\mu,p_\mu)$. The distribution function $f(x,p)$ is defined on the 
future mass shell, satisfying the mass-shell constraint $p^\mu p_\mu = -m^2$. 
For a collisionless Vlasov gas, the distribution function evolves 
according to the Vlasov equation, which can be elegantly expressed 
in terms of the Poisson bracket with the free-particle 
Hamiltonian $\mathcal{H}=\frac12 g^{\mu\nu}(x)p_\mu p_\nu$:
\begin{equation}
\{\mathcal{H},f\} \equiv \frac{\partial \mathcal{H}}{\partial p_\mu}\frac{\partial f}{\partial x^\mu} - \frac{\partial \mathcal{H}}{\partial x^\mu}\frac{\partial f}{\partial p_\mu} = 0.
\end{equation}

The macroscopic properties of the gas are obtained by taking moments of the distribution function in momentum space. The particle current density vector field is defined as:
\begin{equation}
J^\mu(x) = c\int_{p^\mu p_\mu=-m^2} p^\mu f(x,p)\,\text{dvol}_x(p),
\label{eq:J}
\end{equation}
where the Lorentz-invariant volume element on the mass shell is given by
\begin{equation}
\text{dvol}_x(p) := \frac{1}{\sqrt{-g}}\frac{d^3 p_*}{p^0},
\label{eq:dvol}
\end{equation}
with $\sqrt{-g}$ being the square root of the determinant of the metric \cite{Sarbach:2013uba}, and $p_*=(p_i)$ denoting the spatial components of the momentum covector. From the particle current density, the invariant particle number density and the mean four-velocity are respectively defined as
\begin{equation}
n = \frac{1}{c}\sqrt{-J^\mu J_\mu},
\end{equation}
\begin{equation}
u^\mu = J^\mu / n.
\end{equation}

For a distribution function satisfying the Vlasov equation, the current density automatically satisfies the continuity equation,
\begin{equation}
\nabla_\mu J^\mu = 0,
\end{equation}
which ensures particle number conservation. For a steady, spherically symmetric accretion flow, the conserved mass accretion rate can be defined in terms of the areal radius $R$ and the radial component of the particle current, $J^r$:
\begin{equation}
\dot M = m S(R) J^r(R).
\end{equation}

For the Schwarzschild spacetime, the area-radius relation is simply $S(R)=4\pi R^2$. However, in the Simpson-Visser (SV) spacetime, the non-trivial angular part of the metric yields a modified area-radius relation $S(R)=4\pi (R^2+a^2)$, where $a$ is the bounce parameter. This modification directly affects the geometric factor in the mass accretion rate, highlighting the necessity of a detailed relativistic kinetic treatment for the SV spacetime. In the following sections, we will apply this formalism to compute the mass accretion rate for both purely radial infalling particles and infalling particles with angular momentum, considering monoenergetic and Maxwell-Jüttner distributions.

\subsection{Spherical accretion without angular momentum}

As a warmup, we first consider the case where individual particles have zero angular momentum, i.e., purely radial infall. This limit is closely related to the
relativistic radial accretion flows onto static spherically symmetric black
holes studied in Ref.~\cite{Chaverra:2015bya}. In this case, the motion of each particle is completely determined by its conserved energy $E$. To see this, we recall the mass-shell condition $p^\mu p_\mu = -m^2$. For a static, spherically symmetric spacetime, the four-momentum can be expressed in terms of $E$ and the radial momentum $p^r$. Setting the angular momentum to zero, $L=0$, the mass-shell condition yields
\begin{equation}
(p^r)^2 = E^2 - \alpha(r)^2m^2,
\end{equation}
which shows that $p^r$ is determined solely by $E$ and $r$. Consequently, the distribution function, which in general depends on the integrals of motion, reduces to a function of $E$ alone. In the following integrals we use $F(E)$ instead of $f(x,p)$ to denote the distribution function.

With this simplification, the particle current density components in the SV spacetime are given by
\begin{align}
J^r(r) &= \frac{1}{\rho(r)^2}\int_{\alpha(R)m}^{\infty} F(E)\mathrm{d}E,\\
J^t(r) &= \frac{1}{\alpha(r)^2\rho(r)^2}\int_{\alpha(R)m}^{\infty} F(E)\frac{E}
{\sqrt{E^2 - \alpha(r)^2m^2}}\,\mathrm{d}E.
\end{align}
The integration starts at $E_{\text{min}} = m\alpha(R)$, the minimum energy of particles that can cross the injection sphere $R$ inward; this bound applies at all $r \leq R$ because $E$ is conserved.

The mass accretion rate is given by
\begin{equation}
|\dot{M}| = 4\pi m \rho^2 |J^r|
= 4\pi m \int_{\alpha(R)m}^{\infty} F(E)\mathrm{d}E.
\label{eq:Mdot}
\end{equation}
Moreover, at the injection sphere $r=R$, the invariant particle number density is
\begin{equation}
n_{R} = \sqrt{\alpha(R)^2J^t(R)^2 -
\frac{J^r(R)^2}{\alpha(R)^2}}.
\end{equation}

\subsubsection{Monoenergetic Distribution}

For a monoenergetic distribution of the form
\begin{equation}
    F(E) = f_0 \delta(E - E_0)
\end{equation}
with $E_0 = m\alpha(R)\gamma$ and $\gamma = \bigl(1-v_R^2\bigr)^{-1/2}$ 
the Lorentz factor associated with the three-velocity at the injection sphere, 
we obtain
\begin{align}
J^t(R) &= \frac{f_0}{\rho_{R}^2}\frac{\gamma}{\alpha(R)^2\sqrt{\gamma^2 - 1}},\\
J^r(R) &= -\frac{f_0}{\rho_{R}^2},\\
n_{R} &= \frac{f_0}{\rho_{R}^2}\frac{1}{\alpha(R)\sqrt{\gamma^2 - 1}}.
\end{align}
Eliminating $f_0$, the mass accretion rate takes the form
\begin{equation}
\frac{|\dot{M}|}{m n_{R} v_{R}} = 4\pi \rho_{R}^2 \alpha(R) \gamma
\end{equation}

\subsubsection{Maxwell-Jüttner Distribution}

For a Maxwell-Jüttner distribution
\begin{equation}
F(E) = A e^{-\beta E},
\end{equation}
with $\beta = \frac{1}{k_B T}$ and $z = m\beta \alpha(R)$, 
we obtain at the injection sphere
\begin{align}
J^t(R) &= \frac{AmK_1(z)}{\alpha(R)\rho_{R}^2},\\
J^r(R) &= -\frac{A}{\rho_{R}^2}\frac{e^{-z}}{\beta},\\
n_{R}^2 &= \frac{A^2 m^2}{\rho_{R}^4}\left[K_1^2(z) 
- \frac{e^{-2z}}{z^2}\right].
\end{align}
The mass accretion rate becomes
\begin{equation}
\frac{|\dot{M}|}{m n_{R}}
=\frac{4\pi \rho_{R}^2 \alpha(R)}{\sqrt{\big[z e^{z} K_1(z)\big]^2 - 1}}
\end{equation}
where $K_1(z)$ is the modified Bessel function of the second kind of order one. In the nonrelativistic limit $z\gg 1$ where the temperature and particle's velocity is low, this expression reduces to
\begin{equation}
\frac{|\dot{M}|}{m n_{R}} \approx 4\pi \rho_{R}^2 
\alpha(R)\sqrt{\frac{2}{\pi z}}.
\end{equation}

\subsection{Spherical Accretion with Angular Momentum}

To generalize our analysis to the case of accreting particles with angular momentum, it is convenient to work in an orthonormal tetrad frame. The four-momentum can be expanded in this frame as
\begin{equation}
p^\mu \frac{\partial}{\partial x^\mu} = p^0 e_0 + p^1 e_1 + p^2 e_2 + p^3 e_3,
\end{equation}
where the basis vectors are given by
\begin{equation}
e_\alpha = \left( \frac{1}{\alpha(r)} \frac{\partial}{\partial (ct)}, \alpha(r) \frac{\partial}{\partial r}, \frac{1}{\rho} \frac{\partial}{\partial \theta}, \frac{1}{\rho \sin\theta} \frac{\partial}{\partial \phi} \right),
\end{equation}
with $\alpha(r) \equiv \sqrt{1 - 2M/\rho}$ and $\rho \equiv \sqrt{r^2 + a^2}$. The invariant volume element Eq.~(\ref{eq:dvol}) becomes
\begin{equation}
dvol_x(p) = \frac{dp^1 dp^2 dp^3}{\sqrt{m^2 c^2 + (p^1)^2 + (p^2)^2 + (p^3)^2}}.
\label{old_volume}
\end{equation}

Unlike the purely radial case, the particles now possess angular motion in addition to radial momentum. To determine which quantities govern the particle motion, we examine the geodesic equations. Using the mass-shell condition $p^\mu p_\mu = -m^2$ and substituting the conserved energy $E$ and angular momentum $L$, the radial equation of motion reads
\begin{equation}
(p^1)^2 = \frac{E^2}{ \alpha(r)^2} - m^2 - \frac{L^2}{\rho^2}.
\end{equation}
We can thus define the effective potential as
\begin{equation}
V(r) = \alpha(r)^2 \left( m^2 + \frac{L^2}{\rho^2} \right).
\end{equation}
However, $E$ and $L$ alone are not sufficient to fully specify the particle motion. In a spherical coordinate system, the particle's motion can be decomposed into a radial component and a tangential component. To completely describe the tangential motion, we introduce an auxiliary angle $\chi$, which characterizes the orientation of the tangential velocity in the plane orthogonal to the radial direction. In the following, we shall replace the old spatial momentum coordinates $(p^1, p^2, p^3)$ by $(E, L, \chi)$, while keeping the time component of the momentum unchanged:
\begin{equation}
(p^\mu_\pm) = \left( \frac{E}{\alpha(r)}, \pm \frac{\sqrt{E^2 - V(r)}}{\alpha(r)}, \frac{L\cos\chi}{\rho}, \frac{L\sin\chi}{\rho} \right).
\end{equation}
The new form of volume element is obtained by multiplying Eq.~(\ref{old_volume}) by the Jacobian determinant $T$, which is
\begin{equation}
T=\frac{\partial(p^0, E, L, \chi)}{\partial(p^0, p^1, p^2, p^3)},
\end{equation}
and yielding
\begin{equation}
dvol_x(p) = \frac{L dE \, dL \, d\chi}{\rho^2 \sqrt{E^2 - V(r)}}.
\end{equation}

After changing the integration measure from the spatial momentum components to $(E,L,\chi)$, the distribution function is written as $F(E)$. This simplification follows from the fact that, for a steady, spherically symmetric accretion flow, the Vlasov equation requires the distribution function to depend only on the integrals of motion, $E$ and $L$. Since we assume a uniform distribution in the total angular momentum $L$, the dependence on $L$ is trivial, and spherical symmetry further excludes any dependence on the auxiliary angle $\chi$. Consequently, the distribution function depends only on the energy $E$, i.e., $f(x,p)=F(E)$.

\begin{figure}
    \centering
    \includegraphics[width=\columnwidth]{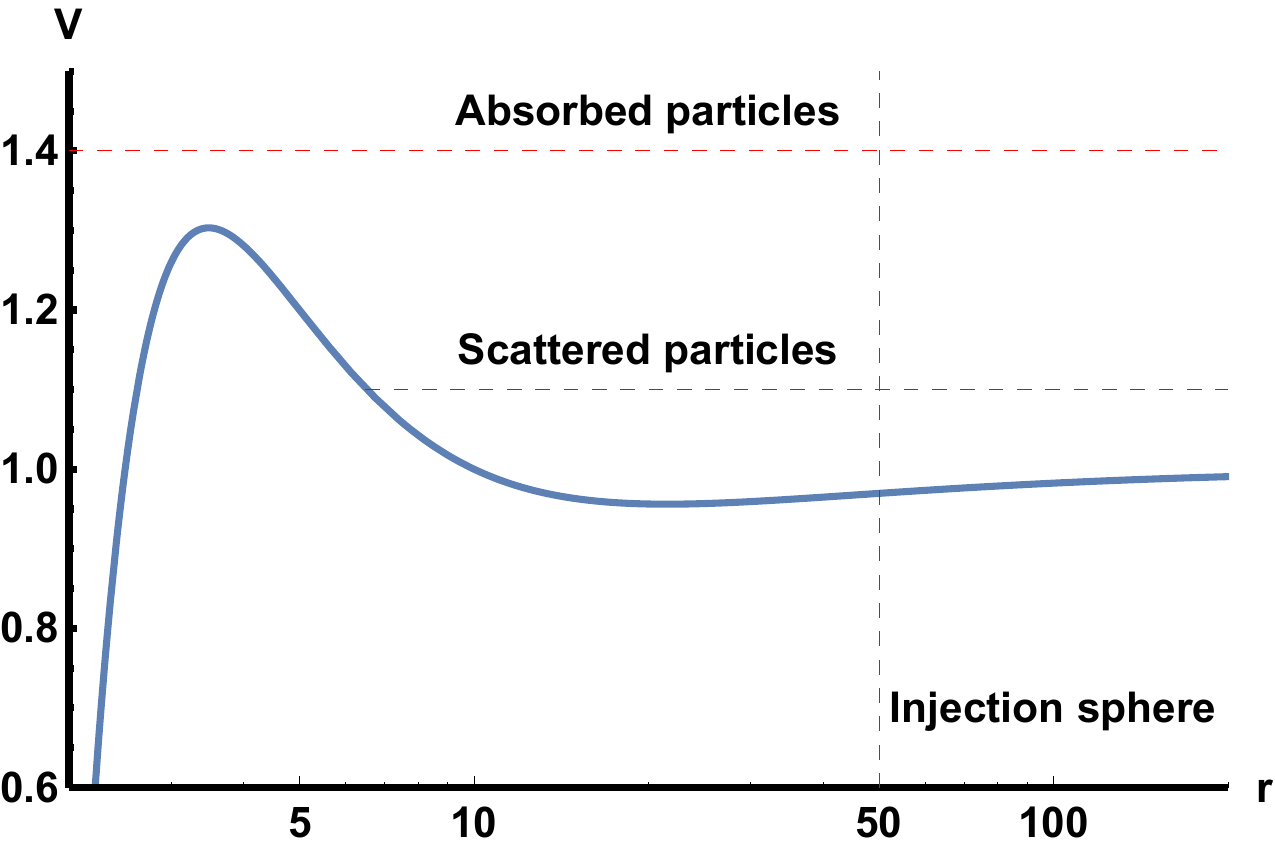}
    \caption{Effective potential $V_L(r)$ for a representative angular momentum $L$, plotted as a function of the radial coordinate $r$ on a logarithmic scale. The curve exhibits a local maximum (centrifugal barrier). The upper horizontal dashed line represents an example energy above the barrier maximum, for which particles are absorbed by the black hole. The lower horizontal dashed line represents an example energy below the barrier maximum but above the potential at the injection sphere, for which particles are reflected by the barrier and thus scattered. The vertical dashed line indicates the injection sphere at radius $R$.}
    \label{fig:EffectivePotential}
\end{figure}

To perform the integration in Eq.~(\ref{eq:J}) explicitly, the integration limits must be specified. Unlike the purely radial case, however, not all particles are absorbed by the black hole. As illustrated in Fig.~\ref{fig:EffectivePotential}, some particles are reflected by the effective potential barrier and are scattered back, never reaching the horizon. An analysis of the effective potential reveals that its shape is controlled by the angular momentum $L$. A larger angular momentum increases the value of the effective potential at a given radius, and only the regions where $E^2 \ge V_L(r)$ correspond to physically allowed particle motion.

For larger angular momentum, the height of the effective potential barrier increases, making it more difficult for particles to fall into the black hole, while higher energy enables them to overcome a higher barrier. Thus, the energy and angular momentum of a particle jointly determine whether it can be absorbed by the black hole.

According to our assumptions, we consider only particles that can reach the injection sphere; therefore, all particles have energies that allow them to overcome the effective potential at the injection sphere. When the angular momentum of a particle is small (i.e.,$ L < L_{\rm ISCO} = \sqrt{12}Mm$), no effective potential barrier exists, and all particles at the injection sphere can fall into the black hole. However, for $L>L_{\mathrm{ISCO}}$, $V_L$ develops a local maximum at $\rho=\rho_-$, defined by
\begin{equation}
    \left.\frac{dV_L}{d\rho}\right|_{\rho_-}=0,\quad
    \left.\frac{d^2V_L}{d\rho^2}\right|_{\rho_-}<0
    \label{eq:rho-}
\end{equation}
As the angular momentum increases, the barrier at $\rho_-$ grows faster than the effective potential at $R$. This leads to a critical situation in which the effective potential at $\rho_-$ equals that at $R$. The energy of a particle that just overcomes both potential values defines the critical energy $E_c(R)$. When the particle energy is below $E_c(R)$, the maximum angular momentum is fixed by
$V(R)|_{L=L_{\max}}=E^2$, because a larger $L$ would give $V(R)>E^2$
and such particles cannot reach $R$. Meanwhile, as the potential at $\rho_-$ grows faster than potential at $R$, for the case below the critical value, the potential $V(\rho_-)$ is not yet large enough, and one always has $V(\rho_-) < V(R)$. Therefore, all particles with $E < E_c(R)$ can fall into the black hole. When the particle energy is above $E_c(R)$, one always has $V(\rho_-) > V(R)$. Only those particles satisfying $E > V(\rho_-)$ can fall into the black hole, hence they must satisfy $L < L_c$, where $E^2 = V(\rho_-)|_{L=L_c}$; those with $E < V(\rho_-)$ are blocked by the barrier, these particles satisfy $L_c < L < L_{\max}$, and thus meet both the conditions of reaching the injection sphere and being scattered by the barrier, as a result, these particles are scattered to infinity instead of being absorbed.

In summary, the particles absorbed by the black hole are those with
\begin{equation}
\begin{cases}
mc^2\alpha(R)\le E < E_c(R) \quad \text{and}\quad 0\le L \le L_{\max}(E,R),\\
E_c(R) < E < \infty \quad \text{and}\quad 0\le L < L_c(E).
\end{cases}
\end{equation}
while the particles scattered by the effective potential barrier are those with
\begin{equation}
E_c(R) \le E < \infty \quad \text{and}\quad L_c(E) < L < L_{\max}(E,R).
\end{equation}

Setting $V_L(R)=E^2$ and solving for $L$ yields the maximum angular momentum
\begin{equation}
L_{\max} = mc\rho_R \sqrt{\frac{E^2}{m^2 \alpha(R)^2} - 1},
\end{equation}
The critical angular momentum $L_c(E)$ is defined by the marginal case $V_L(\rho_-)=E^2$, which gives
\begin{equation}
L_c = \frac{4\sqrt{2}GM m^3}{\sqrt{36m^2 E^2 - 8m^4 - 27E^4 + E(9E^2 - 8m^2)^{3/2}}},
\end{equation}
the critical energy $E_c(R)$ is defined by the condition that the barrier at $\rho_-$ just becomes relevant, i.e. , the maximum angular momentum equals the critical angular momentum:
\begin{equation}
E_c = E_c(R) = m \frac{\rho_R + r_s}{\sqrt{\rho_R(\rho_R + 3r_s)}}.
\end{equation}
The properties of the effective potential and the critical energy
$E_c$ and $L_c$ are discussed in detail in Ref.~\cite{Rioseco:2016jwc}.

The particle‑flow density vector reads
\begin{align}
    \left.J^\sigma_{abs}\right|_{r=R} &=  \int_{mc^2\alpha(R)}^{E_c} 
    \int_{0}^{L_{max}} \int_{0}^{2\pi} \frac{p^\sigma_- F(E)dELdLd\chi}
    {\rho_R^2\sqrt{E^2 - V_L(R)}}\\
    &+\int_{E_c}^{+\infty}\int_{0}^{L_c} \int_{0}^{2\pi}
    \frac{p^\sigma_- F(E)dELdLd\chi}{\rho_R^2\sqrt{E^2 - V_L(R)}},\\
    \left. J^\sigma_{sca} \right|_{r=R} &= \sum_{\pm} \int_{E_c}^{+\infty}
    \int_{L_c}^{L_{max}} \int_{0}^{2\pi} \frac{p^\sigma_\pm F(E)dELdL 
    d\chi}{\rho_R^2\sqrt{E^2 - V_L(R)}},
\end{align}
After further integrating over angular momentum $L$ and angle $\chi$, we find that the nonzero tensor component reads
\begin{align}
J^0_{\mathrm{abs}} \bigg|_{r=R} &= 
\bigg\{
\int_{m \alpha(R)}^{+\infty} 
E\sqrt{E^2 - m^2 \alpha(R)^2} F(E) dE 
\nonumber\\
&\quad - \int_{E_c(R)}^{+\infty} E\sqrt{E^2 - V_c(E,R)} F(E) dE
\bigg\}\frac{2\pi}{\alpha(R)^3 }, \label{J0abs}\\
J^1_{\mathrm{abs}} \bigg|_{r=R} &= 
-\bigg\{
\int_{m \alpha(R)}^{E_c(R)} L_{\text{max}}(E,R)^2 F(E) dE \nonumber\\
&\quad + \int_{E_c(R)}^{+\infty} L_c(E)^2 F(E) dE
\bigg\} 
\frac{\pi}{\rho_R^2 \alpha(R)}, \label{J1abs}\\
J^0_{\mathrm{sca}} \bigg|_{r=R} &= \frac{4\pi}{\alpha(R)^3 }
\int_{E_c(R)}^{+\infty} E\sqrt{E^2 - V_c(E,R)} F(E) dE,\label{J0sca}
\end{align}

\subsubsection{Monoenergetic Model}

For the monoenergetic distribution $F(E)=f_0\delta(E-E_0)$, 
we express the particle energy as $E_0=m\alpha(R)\gamma$, which allows us to replace the integration variable $E$ by the Lorentz factor $\gamma$, we separate the mass accretion rate into two regimes according to whether $\gamma$ lies below or above the critical Lorentz factor $\gamma_c(R)=E_c(R)/\bigl[m\alpha(R)\bigr]$.

For $1<\gamma<\gamma_c$, the mass accretion rate reads
\begin{equation}
    \frac{|\dot{M}|}{m n_R} 
    =4\pi \rho_R^{2}\alpha\sqrt{\frac{\gamma^{2}-1}{3\gamma^{2}+1}}.
    \label{eq:DeltaMdot1}
\end{equation}

For $\gamma>\gamma_c$,the mass accretion rate is as follows
\begin{align}
    \frac{|\dot{M}|}{m n_R} &= 4\pi \rho_R^2 \alpha(R)\frac{h(R,\gamma)}{\sqrt{B(R,\gamma)}} \label{eq:DeltaMdot2}
\end{align}
where
\begin{align}
h(R,\gamma) &:= \left[ \frac{L_c(E_0)}{m\rho_R} \right]^2 \nonumber\\
= \frac{8r_S^2}{\rho_R^2} &\cdot \frac{1}{36\alpha^2\gamma^2 - 8 - 27\alpha^4\gamma^4 + \alpha\gamma\left[9\alpha^2\gamma^2 - 8\right]^{3/2}}. \label{eq:h_def}\\
B(R,\gamma) &=  4\gamma^2\left(\sqrt{\gamma^2-1} + \sqrt{\gamma^2-1-h(R,\gamma)}\right)^2  \nonumber \\
&\quad  - h(R,\gamma)^2 \label{eq:H_def}
\end{align}

After obtaining the piecewise mass accretion rate in the monoenergetic
model, we note that the inflow--outflow model of Shcherbakov and Baganoff
\cite{Shcherbakov:2010yq} predicts a particle velocity at the injection
sphere $R_s=5r_s$ that exceeds the critical value, i.e. $\gamma>\gamma_c$.
Since the SV injection radius $R$ differs from $R_s$ by only about $1\%$,
this condition remains valid for the SV spacetime as well. We therefore focus on the high-energy regime, Eq.~(\ref{eq:DeltaMdot2}).

\begin{figure*}[t]
    \centering
    \begin{minipage}[t]{0.48\linewidth}
        \centering
        \includegraphics[width=\linewidth]{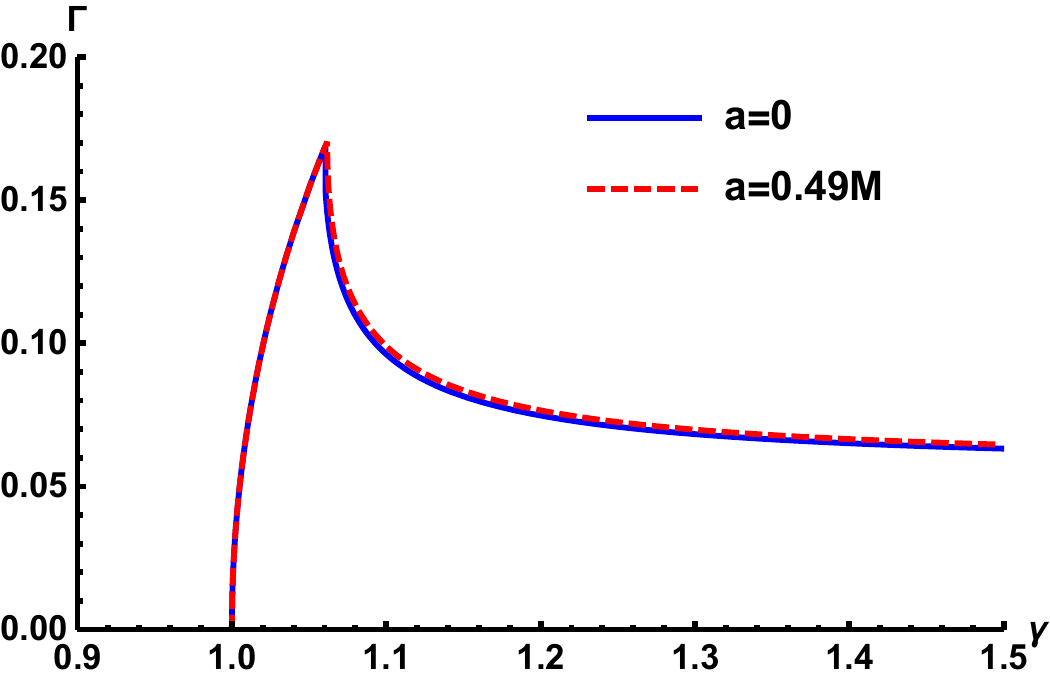}\\[-1mm]
        {\small (a)}
    \end{minipage}\hfill
    \begin{minipage}[t]{0.48\linewidth}
        \centering
        \includegraphics[width=\linewidth]{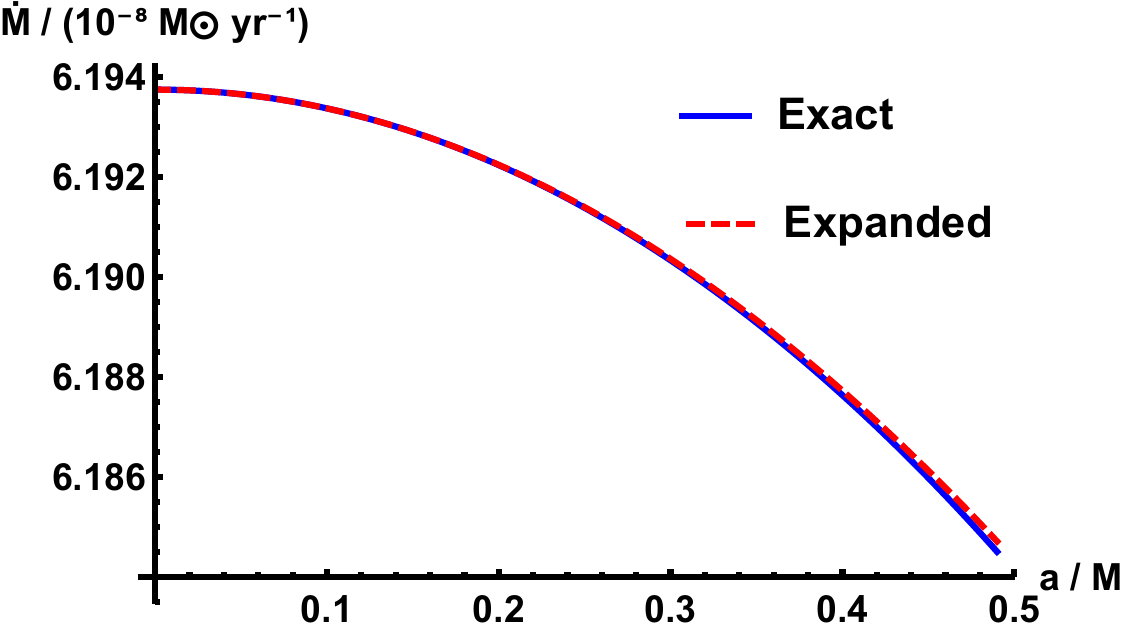}\\[-1mm]
        {\small (b)}
    \end{minipage}
    \caption{Mass accretion rate for the monoenergetic model with angular
        momentum in the Simpson--Visser spacetime.  (a) Dimensionless rate
        $\Gamma_1$ as a function of the Lorentz factor $\gamma$ for $a=0$ and 
        $a=0.49M$.  (b) Dimensional rate $\dot M$ as a function of $a/M$ at 
        $\gamma=1.13259$, corresponding to 
        $v_R=1.4\times10^5\,\text{km}\,\text{s}^{-1}$. The environmental
        parameters are $m=m_p=1.67\times10^{-27}\,\text{kg}$,
        $M=4.3\times10^6\,M_\odot$,
        $n_R=2\times10^6\,\text{cm}^{-3}$, and $R_s=5r_s$.  The exact curve in
        panel (b) is evaluated at the exact SV injection radius $R_{\text{exact}}$
        obtained from $I_L=I_R$, while the expanded curve uses the small-$a$
        expansion in Eq.~(\ref{eq:DeltaMdotExpand2}). The two panels show the same
        rate in dimensionless and dimensional forms, related by
        $\dot M=4\pi\rho_R^2\alpha m n_R\Gamma_1$; they correspond to different
        parameter slices.}
    \label{fig:DeltaMdot}
\end{figure*}

We also evaluate the dimensionless quantity defined by
\begin{equation}
\Gamma_1=
\begin{cases}
\displaystyle\sqrt{\frac{\gamma^2-1}{3\gamma^2+1}}, & 1<\gamma<\gamma_c,\\[6pt]
\displaystyle\frac{h(R,\gamma)}{\sqrt{B(R,\gamma)}}, & \gamma>\gamma_c.
\end{cases}
\label{eq:Gamma_1}
\end{equation}
which characterizes the normalized mass accretion rate for the monoenergetic model. Figure~\ref{fig:DeltaMdot}(a) shows how $\Gamma_1$ varies with the particle Lorentz factor $\gamma$. Starting from the threshold $\gamma \simeq 1.0$, $\Gamma$ rises sharply to a prominent peak and then decreases monotonically for larger Lorentz factors. Within the observationally allowed range of the regularization parameter $a$, the curve corresponding to $a=0.49M$ nearly overlaps with the Schwarzschild ($a=0$) reference solution. At the adopted injection radius $R$, the corrections originating from the Simpson–Visser metric only produce minor quantitative departures relative to the Schwarzschild spacetime for this monoenergetic configuration.

Observations constrain the bounce parameter to $0\le a/M\le 0.49$ \cite{Riaz:2022rlx}. Since the injection radius adopted in our analysis is $R\simeq4.945r_s$, the ratio $a/R$ is at most 0.05. We therefore expand the mass accretion rate around $a=0$ up to second order in the bounce parameter $a$, which allows us to quantify the deviation from the Schwarzschild result in a transparent analytical form. The detailed algebraic derivation is provided in Appendix \ref{app:Mono-Expansion}. The complete expansion of the mass accretion rate in the high-energy regime yields
\begin{equation}
\begin{split}
\frac{|\dot{M}|}{m n_R}
&= 4\pi \left[
\frac{R_s^2 \alpha_0 h_0}{\sqrt{B_0}}
+ \left(
\frac{C' h_0}{\sqrt{B_0}} - \frac{R_s^2 \alpha_0 h_1}{\sqrt{B_0}}
\right. \right. \\
&\quad\quad \left. \left.
- \frac{R_s^2 \alpha_0 h_0 B_1}{2 B_0^{3/2}}
\right) a^2
\right] + \mathcal{O}(a^4).
\label{eq:DeltaMdotExpand2}
\end{split}
\end{equation}
where the coefficients $h_0$, $h_1$, and $B_0$, $B_1$, are intermediate functions defined as in Appendix \ref{app:Mono-Expansion}. 

To illustrate the astrophysical relevance of the monoenergetic model, we apply it 
to the supermassive black hole Sgr A*. Following the two-temperature inflow-
outflow model of Shcherbakov and Baganoff \cite{Shcherbakov:2010yq}, we adopt 
their environmental parameters obtained at the Schwarzschild injection radius 
$R_s=5r_s$, but use the SV injection radius $R$ determined in Sec. \ref{sec:Black 
bounce spacetime}. Specifically, we take $M = 4.3\times 10^6 M_\odot,T=1.2\times 
10^{12}\,\text{K},n_R=2\times 10^6\,\text{cm}^{-3}$. We assume a fully ionized 
hydrogen plasma with $m=m_p$. The three-velocity at the injection sphere is 
estimated by matching $m_p v_R^2$ with $2k_B T$, yielding $v_R\simeq 1.4\times 
10^5\,\text{km}\,\text{s}^{-1}$. For the regularization parameter, we adopt the 
observational upper bound from Riaz et al.\cite{Riaz:2022rlx}, $a/M<0.49$ ($90\%$ 
CL), obtained from X-ray reflection spectroscopy of the Galactic black hole EXO 
1846-031 \cite{Tripathi_2021}.

In Fig.~\ref{fig:DeltaMdot}(b), we plot the mass accretion rate
$\dot M$ as a function of $a/M$ for the monoenergetic model with
angular momentum. The blue solid curve is computed from the exact
expression in Eq.~(\ref{eq:DeltaMdot2}), while the red dashed curve 
is obtained from the small-$a$ expansion in Eq.~(\ref{eq:DeltaMdotExpand2}). 
The two curves are practically indistinguishable over the entire range 
$0\leq a/M\leq 0.49$. The accretion rate decreases from
$6.194\times10^{-8}\,M_\odot\,{\text{yr}}^{-1}$ at $a=0$ to
$6.185\times10^{-8}\,M_\odot\,{\text{yr}}^{-1}$ at $a=0.49M$,
corresponding to a relative change of $-0.149\%$. At large $a$,
the expanded curve lies slightly above the exact one.

To quantify the agreement, we define the relative truncation error
\begin{equation}
    \Delta_2=\frac{|\dot{M}_{\text{exact}}-\dot{M}_{\text{expanded}}|}{\dot{M}_{\text{exact}}}
    \label{eq:error}
\end{equation}

It increases monotonically with $a$ and reaches $3.33\times10^{-5}$ at
the observational upper bound $a/M=0.49$ (see Fig.~\ref{fig:DeltaError} in
Appendix~\ref{app:error}).  This is about 45 times smaller than the
physical change of $0.149\%$, so the truncated expansion reproduces the
exact accretion rate at the $3.3\times10^{-5}$ relative level.  The
residual is consistent with the expected $\mathcal O(a^4)$ truncation
error.

\subsubsection{Maxwell-Jüttner Model}

For the Maxwell-Jüttner distribution $F(E)=Ae^{-\beta E}$, 
we perform a variable substitution $E=m\alpha(R)(1+x/z),
E_c=m\alpha(R)(1+y/z)$ which simplifies the expression of accretion current 
\begin{align}
J^0 &:= \bigl(J^0_{\rm abs}+J^0_{\rm sca}\bigr)\big|_{r=R} = \frac{2\pi A m^3 }{z^{3/2}} e^{-z} I_1(R,z), \\
J^1 &:= J^1_{\rm abs}\big|_{r=R} = -\frac{\pi A m^3 }{z^2} e^{-z} I_2(R,z),
\end{align}
where
\begin{align}
I_1(R,z) &= \int_{0}^{\infty} \left(1+\frac{x}{z}\right) 
\sqrt{2x+\frac{x^2}{z}} e^{-x} dx \nonumber\\
&\quad + \gamma_{\text{c}}(R)^3 e^{-\Lambda(R,z)} \int_{0}^{\infty} 
\left(1+\frac{y}{z}\right) e^{-\gamma_{\text{c}}(R) y} \nonumber\\
\quad \times& \sqrt{ z\left[1-\frac{V_{\text{c}}\left[E_{\text{c}}
(R)\left(1+\frac{y}{z}\right),R\right]}{E_{\text{c}}(R)^2}\right] 
+ 2y + \frac{y^2}{z} } dy,\\
I_2(R,z) &= \int_{0}^{\Lambda(R,z)} \left(2x+\frac{x^2}{z}\right) e^{-x} dx 
+ \frac{\gamma_c(R)}{\rho_R^2} z e^{-\Lambda(R,z)} \nonumber\\
&\quad \times \int_{0}^{\infty} \frac{L_c\left[E_c(R)\left(1+\frac{y}{z}\right)\right]^2}{m^2} e^{-\gamma_c(R) y} dy.
\end{align}
with $\Lambda(R,z) \equiv \bigl[\gamma_c(R)-1\bigr] z$. 
The mass accretion rate takes the form
\begin{equation}
    \frac{|\dot{M}|}{m n_R}= \frac{4\pi \rho_R^2 \alpha(R)}
    {\sqrt{4z\left[\frac{I_1(R,z)}{I_2(R,z)}\right]^2 - 1}}.
    \label{eq:MaxwellMdot}
\end{equation}

\begin{figure*}[t]
    \centering
    \begin{minipage}[t]{0.48\linewidth}
        \centering
        \includegraphics[width=\linewidth]{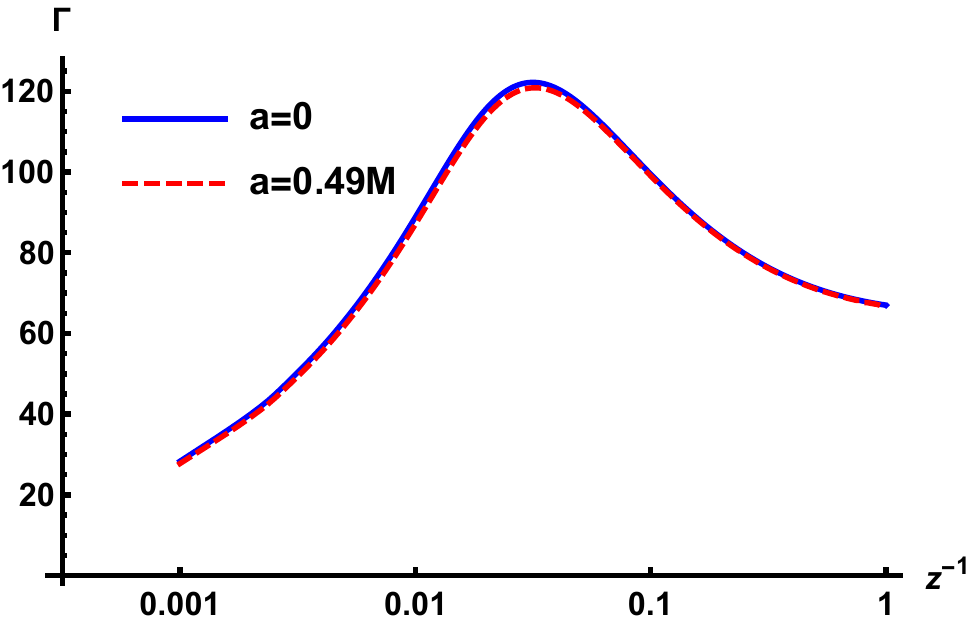}\\[-1mm]
        {\small (a)}
    \end{minipage}\hfill
    \begin{minipage}[t]{0.48\linewidth}
        \centering
        \includegraphics[width=\linewidth]{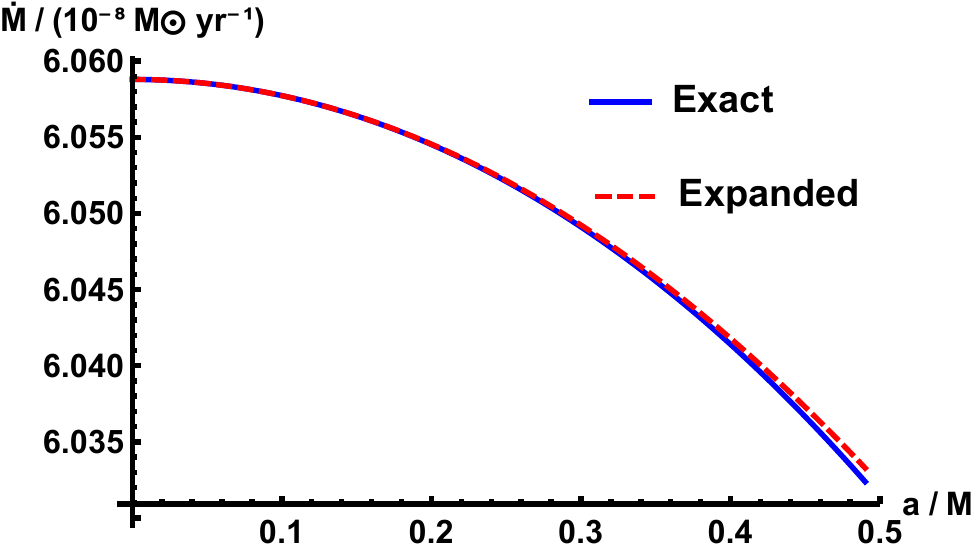}\\[-1mm]
        {\small (b)}
    \end{minipage}
    \caption{Mass accretion rate for the Maxwell--J\"uttner model with
    angular momentum in the Simpson--Visser spacetime.  (a) Dimensionless
    rate
    $\Gamma_2=1/\sqrt{4z[I_1(R,z)/I_2(R,z)]^2-1}$ as a function of
    $z^{-1}=k_BT/(mc^2\alpha(R))$ for $a=0$ and $a=0.49M$.
    (b) Dimensional rate $\dot M$ as a function of $a/M$ at
    $T=1.2\times10^{12}\,\text{K}$, with the exact and expanded curves.
    The remaining parameters are $m=m_p=1.67\times10^{-27}\,\text{kg}$,
    $M=4.3\times10^6\,M_\odot$,
    $n_R=2\times10^6\,\text{cm}^{-3}$, and $R_s=5r_s$.  The exact curve in
    panel (b) is computed from Eq.~(\ref{eq:MaxwellMdot}) with
    $R=R_{\text{exact}}$, while the expanded curve uses the small-$a$
    expansion in Eq.~(\ref{eq:MaxwellMdotExpand}).  The two panels show the
    same rate in dimensionless and dimensional forms, related by
    $\dot M=4\pi\rho_R^2\alpha m n_R\Gamma_2$; they correspond to different
    parameter slices.}
    \label{fig:MaxwellMdot}
\end{figure*}

For the Maxwell–Jüttner distribution, we compute the dimensionless accretion quantity 
\begin{equation}
\Gamma_2 = \frac{1}
    {\sqrt{4z\left[\frac{I_1(R,z)}{I_2(R,z)}\right]^2 - 1}}, 
    \label{eq:Gamma_2}
\end{equation}
where $\Gamma_2$ describes the normalized mass accretion rate for the Maxwell–Jüttner distribution. Figure~\ref{fig:MaxwellMdot}(a) presents the dependence of $\Gamma_2$ on $z^{-1}$. The profile increases at small $z^{-1}$, forms a broad maximum at intermediate values, and gradually declines for larger $z^{-1}$. For observationally consistent values of the regularization parameter $a$, the mass‑accretion‑rate curve for $a=0.49M$ almost coincides with the Schwarzschild result.

We similarly expand the Maxwell–Jüttner mass accretion rate around $a=0$ to second order in the bounce parameter $a$, with details deferred to Appendix \ref{app:Maxwell-Expansion}. The expansion of the mass accretion rate is
\begin{equation}
\frac{|\dot M|}{m n_R}
=
4\pi
\left[
\frac{R_s^2\alpha_0}{\sqrt{W_0}}
+
\left(
\frac{C'}{\sqrt{W_0}}
-\frac{R_s^2\alpha_0W_1}{2W_0^{3/2}}
\right)a^2
\right]
+\mathcal{O}(a^4).
\label{eq:MaxwellMdotExpand}
\end{equation}
where $W_0$, $W_1$ denote two intermediate functions, which are also presented in Appendix \ref{app:Maxwell-Expansion}

For the numerical evaluation of the Maxwell--J\"uttner model, we adopt
the same environmental parameters as in the Schwarzschild case and set
$R_s=5r_s$, namely $M=4.3\times 10^6 M_\odot$, $T=1.2\times 10^{12}\,\text{K}$, 
and $n_R=2\times 10^6\,\text{cm}^{-3}$. The exact curve is computed from Eq.~(\ref{eq:MaxwellMdot}) 
with $R=R_{\text{exact}}$, where $R_{\text{exact}}$ is obtained by solving
$I_L=I_R$. The expanded curve is computed from Eq.~(\ref{eq:MaxwellMdotExpand}), 
in which the $\mathcal{O}(a^2)$ correction to $R(a)$ is already included.

Figure~\ref{fig:MaxwellMdot}(b) shows $\dot M$ as a function of $a/M$ for
the Maxwell--J\"uttner model. As in the monoenergetic case, the two curves
are nearly indistinguishable over the entire range
$0\leq a/M\leq 0.49$, with the expanded curve lying slightly above
the exact one at large $a$. The accretion rate decreases from
$6.059\times10^{-8}\,M_\odot\,{\text{yr}}^{-1}$ at $a=0$ to
$6.032\times10^{-8}\,M_\odot\,{\text{yr}}^{-1}$ at $a=0.49M$,
corresponding to a relative change of $-0.436\%$. The expanded result
at the upper bound is
$6.033\times10^{-8}\,M_\odot\,{\text{yr}}^{-1}$.

To quantify the agreement, we use the relative truncation error defined in
Eq.~(\ref{eq:error}).  It increases monotonically with $a$ and reaches
$1.49\times10^{-4}$ at $a/M=0.49$ (see Fig.~\ref{fig:MaxwellError} in
Appendix~\ref{app:error}).  This is about 29 times smaller than the
physical change of $0.436\%$, so the small-$a$ expansion remains
accurate at the $1.5\times10^{-4}$ relative level over the
observationally allowed range.

\section{Conclusion and Discussion}
\label{sec:Conclusion and Discussion}

\begin{table*}
    \centering
    \begin{tabular*}{\textwidth}{c@{\extracolsep{\fill}}cc}
         && \\ \hline
            Space-time Background & Distribution function & $|\dot{M}|\,[\mathrm{M}_{\odot} \, \mathrm{yr}^{-1}]$
            \\ \hline \hline
            Schwarzschild (Ref.~\cite{Gamboa:2021ytn}) & Maxwell-Jüttner & $~5.77\times10^{-8}$\\
            Simpson-Visser (a=0, this work) & Monoenergetic & $~6.19\times10^{-8}$\\
            Simpson-Visser (a=0.49M, this work) & Monoenergetic & $~6.18\times10^{-8}$\\
            Simpson-Visser (a=0, this work) & Maxwell-Jüttner & $~6.06\times10^{-8}$\\
            Simpson-Visser (a=0.49M, this work) & Maxwell-Jüttner & $~6.03\times10^{-8}$\\
    \hline
    \end{tabular*}
    \caption{Mass accretion rate inferred for Sgr A* at $R$ for the models studied in this work and results from the literature. We have used $m=m_p=1.67\times10^{-27}\,\text{kg}$,\,$M=4.3\times10^6
    \,\text{M}_{\odot}$,\,$T=1.2\times10^{12}\,\text{K}$,\,$v_R\sim 1.4\times10^5\,\text{km}\,\text{s}^{-1}$ and $n_R=2\times10^{6}\,\text{cm}^{-3}$.}
    \label{tab:applications}
\end{table*}

In this work, we have extended the analysis of Gamboa et al.~\cite{Gamboa:2021ytn}
to the Simpson--Visser (SV) black-bounce spacetime. We derived the mass
accretion rate for purely radial infalling and exact integral representations
for infalling with angular momentum, considering both monoenergetic and
Maxwell--J\"uttner distributions. These expressions reduce correctly to
the Schwarzschild results in the limit $a\to0$.

The main conclusion in this work is that the black bounce parameter $a$ can reduce the mass accretion rate compared to the Schwarzschild case. Because the shadows of the SV and Schwarzschild black holes are degenerate, the accretion rate can provide a useful complementary
probe of the spacetime geometry. We further derived semi-analytic expansions of the mass accretion rate through $\mathcal{O}(a^2)$ around the Schwarzschild limit. Over the
observationally allowed range $0\leq a/M\leq0.49$, the expansions
reproduce the exact rates with maximum relative errors of
$3.33\times10^{-5}$ for the monoenergetic model and
$1.49\times10^{-4}$ for the Maxwell--J\"uttner model. They therefore
provide an accurate and efficient approximation without requiring the
exact integral equation to be solved at every value of $a$.

Applying our results to Sgr~A$^*$ with the environmental parameters of
Shcherbakov and Baganoff~\cite{Shcherbakov:2010yq}, we determined the SV 
injection radius $R$ by fixing the proper radial distance from the horizon. 
At $a/M=0.49$, the accretion rate is reduced by $0.149\%$ in the
monoenergetic case and by $0.436\%$ in the Maxwell--J\"uttner case,
relative to $a=0$. The resulting mass accretion rates are summarized in Table~\ref{tab:applications}. The absolute normalization at $a=0$
differs by about $5\%$ from that of Ref.~\cite{Gamboa:2021ytn}. 
This overall offset between two numerical evaluations of 
the same phase-space integrals, most likely associated with 
the integrable endpoint at $E_c$ rather than with the
$a$-expansion, is independent of $a$ and cancels in the relative change
with respect to $a=0$; it therefore does not affect our main conclusions.

Even after accounting for the shift of the injection sphere, the SV
regularization has only a minor effect on the mass accretion rate for the
parameter range allowed by current observations. This supports the
interpretation that the reduction of the accretion rate due to angular
momentum, as proposed by Gamboa et al.~\cite{Gamboa:2021ytn}, remains a viable
explanation for the low luminosity of sources such as Sgr~A$^*$ in the
SV scenario. 

Several directions remain for future work. A natural extension is to
rotating SV or Kerr-like black-bounce spacetimes \cite{Mazza:2021rgq}, where 
frame dragging and non-spherical accretion invalidate the spherically symmetric 
treatment adopted here. It would also be important to consider more general
distribution functions and angular-momentum distributions, including
anisotropic, power-law, and Fermi--Dirac distributions, as well as the
coupling of the kinetic gas to magnetic fields and radiative transfer.
Comparisons with hydrodynamic and general-relativistic magnetohydrodynamic
simulations would help clarify the extent to which the present kinetic
results can be applied to realistic accretion flows. Finally, extending
the expansion beyond second order, or performing exact numerical
calculations for larger values of $a$---including the near-extremal limit
$a\to2M$ and the wormhole regime $a>2M$---would be valuable for testing
the theoretical robustness of the present results. Although these regimes
lie outside the current observational bound $a/M\le0.49$, they are of
intrinsic theoretical interest and may become relevant if future
observations relax or extend this bound.

\begin{acknowledgments}
This work is supported by the National Natural Science Foundation of China (NSFC) under Grant Nos.12405066, 12175105 and NUAA funding (No. 1018-ILF26028). YSA is also supported by the Natural Science Foundation of Jiangsu Province under Grant No. BK20241376 and Fundamental Research Funds for the Central Universities. 
\end{acknowledgments}

\appendix

\section{Determination of the SV Injection Radius}
\label{app:injection_sphere}

To compare the accretion flow in the Schwarzschild and Simpson--Visser (SV) spacetimes on the same physical scale, we fix the proper radial distance from the horizon to the injection sphere. In the Schwarzschild spacetime, the proper distance from the horizon $r_s = 2M$ to a reference radius $R_s = 5r_s$ is
\begin{equation}
I_R = \int_{r_s}^{R_s} \frac{dr}{\sqrt{1 - r_s/r}}.
\end{equation}
In the SV spacetime, the corresponding proper distance from the horizon $r_H = \sqrt{r_s^2 - a^2}$ to a radius $R$ is
\begin{equation}
I_L = \int_{r_H}^{R} \frac{dr}{\sqrt{1 - r_s/\sqrt{r^2+a^2}}},
\end{equation}
where $a$ is the bounce parameter. Requiring $I_R = I_L$ determines the SV injection radius $R$.

Introduce the change of variable
\begin{equation}
\rho = \sqrt{r^2 + a^2},\quad r = \sqrt{\rho^2 - a^2},\quad dr = \frac{d\rho}{\sqrt{1 - a^2/\rho^2}}.
\end{equation}
The lower limit is $r = r_H \implies \rho = r_s$, while the upper limit is denoted by $\rho_{SV} = \sqrt{R^2 + a^2}$. Thus
\begin{equation}
I_L = \int_{r_s}^{\rho_{SV}} \frac{d\rho}{\sqrt{1 - r_s/\rho}\sqrt{1 - a^2/\rho^2}}.
\end{equation}
For $a \ll r_s$ and $\rho \ge r_s$, we expand the second factor:
\begin{equation}
\frac{1}{\sqrt{1 - a^2/\rho^2}} = 1 + \frac{a^2}{2\rho^2} + \mathcal{O}\left(\frac{a^4}{\rho^4}\right).
\end{equation}
Define $g(x) = \int_{r_s}^{x} \frac{d\rho}{\sqrt{1 - r_s/\rho}}$, such that $I_R = g(R_s)$. Substituting the expansion into the integral for $I_L$ gives
\begin{equation}
I_L = g(\rho_{SV}) + \frac{a^2}{2}\int_{r_s}^{\rho_{SV}} \frac{d\rho}{\rho^2\sqrt{1 - r_s/\rho}} + \mathcal{O}\left(\frac{a^4}{r_s^3}\right).
\end{equation}
Using $\frac{d}{d\rho}\sqrt{1 - r_s/\rho} = \frac{r_s}{2\rho^2\sqrt{1 - r_s/\rho}}$, the integral can be evaluated exactly. Therefore, the condition $I_L = I_R$ becomes
\begin{equation}
g(\rho_{SV}) + \frac{a^2}{r_s}\sqrt{1 - \frac{r_s}{\rho_{SV}}} = g(R_s) + \mathcal{O}\left(\frac{a^4}{r_s^3}\right).
\end{equation}
Since $\rho_{SV} = R_s$ when $a=0$, we write $\rho_{SV} = R_s + a^2 c + \mathcal{O}(a^4)$. Using $g'(x)|_{x=R_s} = 1/\sqrt{1 - r_s/R_s}$, we obtain
\begin{equation}
g(\rho_{SV}) = g(R_s) + \frac{a^2 c}{\sqrt{1 - r_s/R_s}} + \mathcal{O}\left(a^4\right).
\end{equation}
The second term on the left-hand side already carries a factor $a^2$, so we may set $\rho_{SV} = R_s$ inside it. Comparing the $a^2$ coefficients yields
\begin{equation}
\frac{c}{\sqrt{1 - r_s/R_s}} + \frac{1}{r_s}\sqrt{1 - \frac{r_s}{R_s}} = 0,
\end{equation}
hence $c = -\frac{1}{r_s}\left(1 - \frac{r_s}{R_s}\right)$. Thus $\rho_{SV} = R_s - \frac{a^2}{r_s} + \frac{a^2}{R_s} + \mathcal{O}(a^4)$.

Since $\rho_{SV} = \sqrt{R^2 + a^2}$, we have $R = \sqrt{\rho_{SV}^2 - a^2} = \rho_{SV} - \frac{a^2}{2\rho_{SV}} + \mathcal{O}(a^4)$. In the $a^2$ term we may replace $\rho_{SV}$ by $R_s$, yielding the second-order expansion:
\begin{equation}
R^{(2)} = R_s - \frac{a^2}{r_s} + \frac{a^2}{2R_s} + \mathcal{O}\left(\frac{a^4}{r_s^3}\right).
\end{equation}

Extending the expansion to the fourth order yields
\begin{equation}
R^{(4)} = R_s + c_2 a^2 + c_4 a^4 + \mathcal{O}(a^6),
\end{equation}

\begin{figure}
    \centering
    \includegraphics[width=\columnwidth]{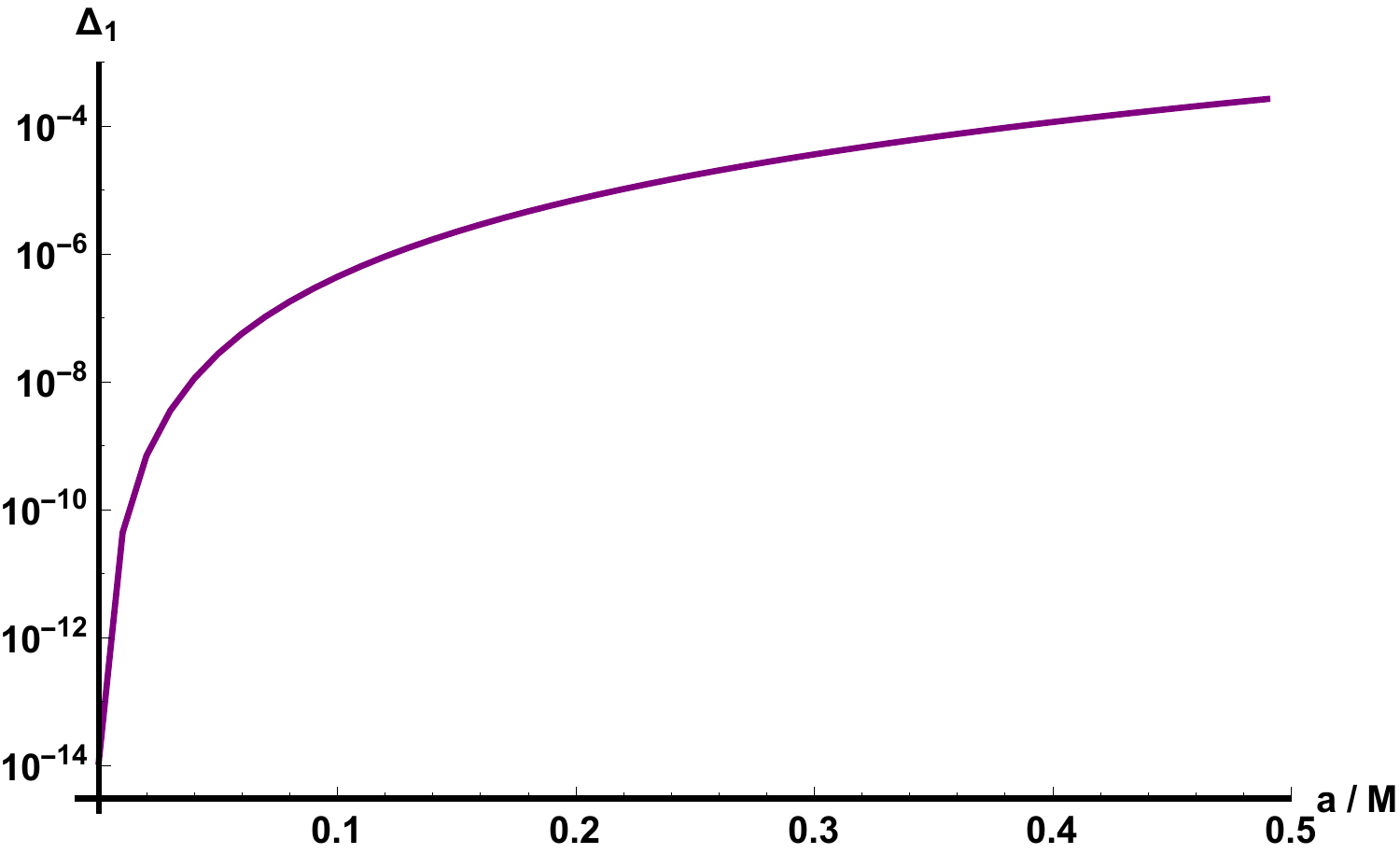}
    \caption{Relative error $\Delta_1 = |R_{\rm exact} - R^{(2)}| / R_{\rm exact}$ of the second-order expansion for the SV injection radius as a function of $a/M$. The vertical axis is logarithmic. The error remains below $2.65 \times 10^{-4}$ for all $a/M \le 0.49$.}
    \label{fig:Rerror}
\end{figure}

where the coefficients are
\begin{align}
c_2 &= -\frac{1}{r_s} + \frac{1}{2R_s}, \\
c_4 &= -\frac{2}{5r_s^3} + \frac{1}{5r_s^2 R_s} + \frac{3}{10r_s R_s^2} - \frac{9}{40R_s^3}.
\end{align}

Numerically solving the exact equation $I_R = I_L$ at the maximum allowed value $a = 0.49M$ (with $R_s = 10M$), we find $R_{\rm exact} \approx 9.8893M$. The second-order expansion gives $R^{(2)} \approx 9.8920M$, while the fourth-order expansion gives $R^{(4)} \approx 9.8894M$. The fourth-order correction accounts for only $\sim 2.3\%$ of the second-order shift, and its inclusion reduces the relative error from $\sim 10^{-4}$ to $\sim 10^{-5}$. Thus, the higher-order terms are well-behaved and negligible, justifying the truncation at second order in the main text.

The relative error $\Delta_1$ defined in Eq.~(\ref{eq:DeltaR}) is shown in Fig.~\ref{fig:Rerror}; its
maximum over $0\le a/M\le0.49$ is $2.65\times10^{-4}$.

\section{Detailed derivation of the mass accretion rate expansion}
\label{app:Expansion}

\subsection{Monoenergetic distribution}
\label{app:Mono-Expansion}

We begin by expanding the fundamental quantities in powers of $a^2$, which are valid for both monoenergetic and Maxwellian distributions. The actual position of the injection sphere in the SV spacetime reads
\begin{equation}
R(a) = R_s + R_1 a^2 + \mathcal{O}(a^4), \quad R_1 = \frac{1}{2R_s} - \frac{1}{r_s}.
\end{equation}
The transformed radial coordinate is
\begin{equation}
\rho_R = \sqrt{R^2+a^2} = R_s + \rho_1 a^2 + \mathcal{O}(a^4),
\end{equation}
with $\rho_1 = \frac{1}{R_s} - \frac{1}{r_s}$. The lapse function is given by
\begin{equation}
\alpha(R) = \sqrt{1-\frac{r_s}{\rho_R}} = \alpha_0 + \alpha_1 a^2 + \mathcal{O}(a^4),
\end{equation}
where
\begin{equation}
\alpha_0 = \sqrt{1-\frac{r_s}{R_s}}, \quad
\alpha_1 = \frac{r_s-R_s}{2 R_s^3 \alpha_0}.
\end{equation}
The combination term $\rho_R^2\alpha(R)$ expands to
\begin{equation}
\rho_R^2\alpha(R) = R_s^2 \alpha_0 + C' a^2 + \mathcal{O}(a^4),
\label{eq:app_rho2alpha}
\end{equation}
with the coefficient $C' = 2R_s\left(\frac{1}{R_s}-\frac{1}{r_s}\right)\alpha_0 + R_s^2 \alpha_1$.

For the monoenergetic distribution, we separate the expansion into the low-energy and high-energy regimes. In the low-energy regime $1<\gamma<\gamma_c$, substituting Eq.~(\ref{eq:app_rho2alpha}) directly into the accretion rate formula, we obtain
\begin{equation}
\frac{|\dot{M}|}{m n_R} = 4\pi \left[R_s^2 \alpha_0 + C' a^2\right] \sqrt{\frac{\gamma^2-1}{3\gamma^2+1}} + \mathcal{O}(a^4).
\label{eq:app_DeltaMdot_low}
\end{equation}

In the high-energy regime $\gamma>\gamma_c$, the mass accretion rate takes the form
\begin{equation}
\frac{|\dot{M}|}{m n_R} = 4\pi \rho_R^2 \alpha(R) \frac{h}{\sqrt{B}},
\end{equation}
where the polynomial $A$, the function $h$, and the denominator function $B$ are defined as
\begin{align}
A &= 36\alpha^2\gamma^2 - 8 - 27\alpha^4\gamma^4 + \alpha\gamma\left(9\alpha^2\gamma^2 - 8\right)^{3/2}, \label{eq:A_def}\\
h &= \frac{8 r_s^2}{\rho_R^2 A},\\
B &= 4\gamma^2\left(\sqrt{\gamma^2-1} + \sqrt{\gamma^2-1-h}\right)^2 - h^2.
\end{align}

We expand each quantity systematically. For polynomial $A$, writing $\alpha=\alpha_0+\alpha_1 a^2$, we have $A = A_0 + A_1 a^2 + \mathcal{O}(a^4)$, with
\begin{align}
A_0 &= 36\alpha_0^2\gamma^2 - 8 - 27\alpha_0^4\gamma^4 +
\alpha_0\gamma\left(9\alpha_0^2\gamma^2 - 8\right)^{3/2},\\
A_1 &= \frac{2(r_s-R_s)}{r_s} \biggl[
\frac{18 r_s \gamma^2}{R_s^3}
- \frac{27 \alpha_0^2 r_s \gamma^4}{R_s^3}  \nonumber\\
&\quad + \frac{r_s \gamma}{4 R_s^3 \alpha_0}\left(9\alpha_0^2\gamma^2 - 8\right)^{3/2}
+ \frac{27 r_s \alpha_0 \gamma^3}{4 R_s^3}\sqrt{9\alpha_0^2\gamma^2 - 8}
 \biggr].
\end{align}

For function $h$, using the expansions $1/\rho_R^2 = 1/R_s^2 - 2\rho_1 a^2/R_s^3 + \mathcal{O}(a^4)$ and $1/A = 1/A_0 - A_1 a^2/A_0^2 + \mathcal{O}(a^4)$, we find $h = h_0 - h_1 a^2 + \mathcal{O}(a^4)$, where
\begin{align}
h_0 &= \frac{8 r_s^2}{R_s^2 A_0}, \label{h0}\\
h_1 &= 8 r_s^2\left( \frac{A_1}{R_s^2 A_0^2} + \frac{2\rho_1}{R_s^3 A_0} \right). \label{h1}
\end{align}

For function $B$, let $B' = \sqrt{\gamma^2-1-h_0}$. The square root term expands as
\begin{equation}
\sqrt{\gamma^2-1-h} = B' + \frac{h_1}{2B'}a^2 + \mathcal{O}(a^4).
\end{equation}
Thus $B = B_0 + B_1 a^2 + \mathcal{O}(a^4)$ with
\begin{align}
B_0 &= 4\gamma^2\left(\sqrt{\gamma^2-1} + B'\right)^2 - h_0^2, \label{B0}\\
B_1 &= 4\gamma^2 \frac{h_1\left(\sqrt{\gamma^2-1} + B'\right)}{B'} + 2h_0 h_1. \label{B1}
\end{align}

The combination $h/\sqrt{B}$ expands to
\begin{equation}
\frac{h}{\sqrt{B}} = \frac{h_0}{\sqrt{B_0}} - \left( \frac{h_1}{\sqrt{B_0}} + \frac{h_0 B_1}{2 B_0^{3/2}} \right) a^2 + \mathcal{O}(a^4).
\end{equation}

Combining all terms, the complete expansion of the mass accretion rate in the high-energy regime is
\begin{equation}
\begin{split}
\frac{|\dot{M}|}{m n_R}
&= 4\pi \left[
\frac{R_s^2 \alpha_0 h_0}{\sqrt{B_0}}
+ \left(
\frac{C' h_0}{\sqrt{B_0}} - \frac{R_s^2 \alpha_0 h_1}{\sqrt{B_0}}
\right. \right. \\
&\quad\quad \left. \left.
- \frac{R_s^2 \alpha_0 h_0 B_1}{2 B_0^{3/2}}
\right) a^2
\right] + \mathcal{O}(a^4).
\label{eq:app_DeltaMdot_high}
\end{split}
\end{equation}

\subsection{Maxwell–Jüttner distribution}
\label{app:Maxwell-Expansion}

Expanding the injection radius and the lapse function through $\mathcal{O}(a^2)$ gives
\begin{equation}
    R=R_s+R_1a^2+\mathcal{O}(a^4),
\end{equation}
\begin{equation}
    \rho_R=R_s+\rho_1a^2+\mathcal{O}(a^4),
\end{equation}
\begin{equation}
    \alpha(R)=\alpha_0+\alpha_1a^2+\mathcal{O}(a^4),
\end{equation}
\begin{equation}
    \rho_R^2\alpha(R)=R_s^2\alpha_0+C'a^2+\mathcal{O}(a^4),
\end{equation}
where $R_1$, $\rho_1$, $\alpha_1$, and
\begin{equation}
C'=2R_s\rho_1\alpha_0+R_s^2\alpha_1
\end{equation}
are defined in Appendix B1. For the mass accretion rate expanded through
$\mathcal{O}(a^2)$, only the $\mathcal{O}(a^2)$ term of $R(a)$ is required; the $\mathcal{O}(a^4)$ term
in $R(a)$ contributes at $\mathcal{O}(a^4)$.

The critical energy is
\begin{equation}
E_c(R)=m\,g(\rho_R),
\qquad
g(\rho)=\frac{\rho+r_s}{\sqrt{\rho(\rho+3r_s)}}.
\end{equation}
Hence
\begin{equation}
E_c=E_{c0}+E_{c1}a^2+\mathcal{O}(a^4),
\end{equation}
with
\begin{equation}
E_{c0}=m\,g_0,
\qquad
E_{c1}=m\,g'_0\,\rho_1,
\end{equation}
where $g_0\equiv g(R_s)$, $g'_0\equiv g'(R_s)$, and
\begin{equation}
g'(\rho)
=
g(\rho)\left[
\frac{1}{\rho+r_s}
-\frac{1}{2}\left(\frac{1}{\rho}+\frac{1}{\rho+3r_s}\right)
\right].
\end{equation}
The critical Lorentz factor is
\begin{equation}
\gamma_c(R)=\frac{E_c(R)}{m\alpha(R)}
=\gamma_0+\gamma_1a^2+\mathcal{O}(a^4),
\end{equation}
with
\begin{equation}
\gamma_0=\frac{g_0}{\alpha_0},
\qquad
\gamma_1=
\frac{g'_0\rho_1\alpha_0-g_0\alpha_1}{\alpha_0^2}.
\end{equation}
For the temperature variable,
\begin{equation}
z=\frac{m\alpha(R)}{k_BT}
=z_0+z_1a^2+\mathcal{O}(a^4),
\end{equation}
where
\begin{equation}
z_0=\frac{m\alpha_0}{k_BT},
\qquad
z_1=\frac{m\alpha_1}{k_BT}.
\end{equation}
Similarly,
\begin{equation}
\Lambda(R,z)\equiv[\gamma_c(R)-1]z
=\Lambda_0+\Lambda_1a^2+\mathcal{O}(a^4),
\end{equation}
with
\begin{equation}
\Lambda_0=(\gamma_0-1)z_0,
\qquad
\Lambda_1=\gamma_1z_0+(\gamma_0-1)z_1.
\end{equation}

The critical angular momentum is
\begin{equation}
L_c(E)=\frac{4\sqrt{2}\,M m}{\sqrt{D(E)}},
\end{equation}
where
\begin{equation}
D(E)
=
36\frac{E^2}{m^2}-8-27\frac{E^4}{m^4}
+\frac{E}{m}\left(9\frac{E^2}{m^2}-8\right)^{3/2}
\end{equation}
is dimensionless. Its derivative is
\begin{equation}
D'(E)
=
\frac{72E}{m^2}-\frac{108E^3}{m^4}
+\frac{\sqrt{9E^2-8m^2}\left(36E^2-8m^2\right)}{m^4},
\end{equation}
so that
\begin{equation}
L_c'(E)
=
-2\sqrt{2}\,M m\,D(E)^{-3/2}D'(E).
\end{equation}

For the argument entering the Maxwell--J\"uttner integrals,
\begin{equation}
L_c\!\left[E_c(R)\left(1+\frac{y}{z}\right)\right]
=L_0(y)+L_1(y)a^2+\mathcal{O}(a^4),
\end{equation}
where
\begin{equation}
L_0(y)=L_c\!\left(E_{c0}\left(1+\frac{y}{z_0}\right)\right),
\end{equation}
and
\begin{align}
L_1(y)
&=
L_c'\!\left(E_{c0}\left(1+\frac{y}{z_0}\right)\right)\nonumber\\
 &\quad \times\left[
E_{c1}\left(1+\frac{y}{z_0}\right)
-\frac{E_{c0}\,y\,z_1}{z_0^2}
\right].
\end{align}

The effective potential at the injection sphere is
\begin{equation}
V_c=\alpha(R)^2
\left(
m^2+\frac{L_c^2}{\rho_R^2}
\right).
\end{equation}
Writing
\begin{equation}
V_c=V_0(y)+V_1(y)a^2+\mathcal{O}(a^4),
\end{equation}
we obtain
\begin{equation}
V_0(y)=
\alpha_0^2
\left(
m^2+\frac{L_0^2}{R_s^2}
\right),
\end{equation}
and
\begin{equation}
V_1(y)=
2\alpha_0\alpha_1
\left(
m^2+\frac{L_0^2}{R_s^2}
\right)
+\frac{2\alpha_0^2L_0L_1}{R_s^2}
-\frac{2\alpha_0^2L_0^2\rho_1}{R_s^3}.
\label{eq:V1}
\end{equation}
The last term contains $\rho_1$, because $V_c$ depends on the areal
radius $\rho_R=R_s+\rho_1a^2$, not on $R$ alone. At $y=0$,
$V_c=E_c^2$, so Eq.~\eqref{eq:V1} satisfies
\begin{equation}
V_0(0)=E_{c0}^2,
\qquad
V_1(0)=2E_{c0}E_{c1}.
\end{equation}

We now expand
\begin{equation}
I_1=G+H,
\end{equation}
where
\begin{equation}
G=
\int_0^\infty
\left(1+\frac{x}{z}\right)
\sqrt{2x+\frac{x^2}{z}}\,e^{-x}\,dx,
\end{equation}
and
\begin{align}
H=
\gamma_c^3e^{-\Lambda}
&\int_0^\infty
\left(1+\frac{y}{z}\right)e^{-\gamma_c y}\nonumber\\
&\times \sqrt{
z\left(1-\frac{V_c}{E_c^2}\right)
+2y+\frac{y^2}{z}
}\,dy.
\end{align}
Define
\begin{equation}
P(y)=
z\left(1-\frac{V_c}{E_c^2}\right)
+2y+\frac{y^2}{z}
=
P_0(y)+P_1(y)a^2+\mathcal{O}(a^4),
\end{equation}
with
\begin{equation}
P_0(y)=
z_0\left(1-\frac{V_0}{E_{c0}^2}\right)
+2y+\frac{y^2}{z_0},
\end{equation}
and
\begin{equation}
P_1(y)=
z_1\left(1-\frac{V_0}{E_{c0}^2}\right)
-z_0\left(
\frac{V_1}{E_{c0}^2}
-\frac{2V_0E_{c1}}{E_{c0}^3}
\right)
-\frac{y^2z_1}{z_0^2}.
\end{equation}

The expansion of $G$ is
\begin{equation}
G=G_0+G_1a^2+\mathcal{O}(a^4),
\end{equation}
with
\begin{equation}
G_0=
\int_0^\infty
\left(1+\frac{x}{z_0}\right)
\sqrt{2x+\frac{x^2}{z_0}}\,e^{-x}\,dx,
\end{equation}
and
\begin{equation}
\begin{aligned}
G_1
=
z_1\int_0^\infty
\Bigg[
&-\frac{x}{z_0^2}\sqrt{2x+\frac{x^2}{z_0}}\\
&-\frac{x^2}{2z_0^2}
\frac{1+x/z_0}{\sqrt{2x+x^2/z_0}}
\Bigg]e^{-x}\,dx.
\end{aligned}
\end{equation}
Similarly,
\begin{equation}
H=H_0+H_1a^2+\mathcal{O}(a^4),
\end{equation}
where
\begin{equation}
H_0=
\gamma_0^3e^{-\Lambda_0}
\int_0^\infty
\left(1+\frac{y}{z_0}\right)
e^{-\gamma_0 y}\sqrt{P_0}\,dy,
\end{equation}
and
\begin{equation}
\begin{aligned}
H_1
=
e^{-\Lambda_0}
&\int_0^\infty
\Bigg\{
\left(3\gamma_0^2\gamma_1-\gamma_0^3\Lambda_1\right)
\left(1+\frac{y}{z_0}\right)
e^{-\gamma_0 y}\sqrt{P_0}\\
&+\gamma_0^3
\left[
-\gamma_1y\left(1+\frac{y}{z_0}\right)
-\frac{yz_1}{z_0^2}
\right]
e^{-\gamma_0 y}\sqrt{P_0}\\
&+\gamma_0^3
\left(1+\frac{y}{z_0}\right)
e^{-\gamma_0 y}
\frac{P_1}{2\sqrt{P_0}}
\Bigg\}\,dy.
\end{aligned}
\end{equation}
Therefore,
\begin{align}
I_1&=I_{10}+I_{11}a^2+\mathcal{O}(a^4),\\
I_{10}&=G_0+H_0,\\
I_{11}&=G_1+H_1.
\end{align}

For the second integral,
\begin{equation}
I_2=K+N,
\end{equation}
where
\begin{equation}
K=
\int_0^\Lambda
\left(2x+\frac{x^2}{z}\right)e^{-x}\,dx,
\end{equation}
and
\begin{equation}
N=
\frac{\gamma_c z}{\rho_R^2}e^{-\Lambda}
\int_0^\infty
\frac{L_c\!\left[E_c\left(1+\frac{y}{z}\right)\right]^2}{m^2}
e^{-\gamma_c y}\,dy.
\end{equation}
The expansion of $K$ is
\begin{equation}
K=K_0+K_1a^2+\mathcal{O}(a^4),
\end{equation}
with
\begin{equation}
K_0=
\int_0^{\Lambda_0}
\left(2x+\frac{x^2}{z_0}\right)e^{-x}\,dx,
\end{equation}
and
\begin{equation}
K_1=
\Lambda_1
\left(
2\Lambda_0+\frac{\Lambda_0^2}{z_0}
\right)e^{-\Lambda_0}
-z_1
\int_0^{\Lambda_0}
\frac{x^2}{z_0^2}e^{-x}\,dx.
\end{equation}
For $N$, define
\begin{equation}
U(y)\equiv\frac{L_c^2}{m^2}
=U_0(y)+U_1(y)a^2+\mathcal{O}(a^4),
\end{equation}
with
\begin{equation}
U_0=\frac{L_0^2}{m^2},
\qquad
U_1=\frac{2L_0L_1}{m^2}.
\end{equation}
Then
\begin{equation}
N=N_0+N_1a^2+\mathcal{O}(a^4),
\end{equation}
where
\begin{equation}
N_0=
\frac{\gamma_0z_0}{R_s^2}e^{-\Lambda_0}
\int_0^\infty
U_0\,e^{-\gamma_0 y}\,dy,
\end{equation}
and
\begin{equation}
\begin{aligned}
N_1
=
e^{-\Lambda_0}
\Bigg[
&\frac{
\gamma_1z_0+\gamma_0z_1-\gamma_0z_0\Lambda_1
-2\gamma_0z_0\rho_1/R_s
}{R_s^2}\\
\times &\int_0^\infty
U_0\,e^{-\gamma_0 y}\,dy\\
&+\frac{\gamma_0z_0}{R_s^2}
\int_0^\infty
\left(
U_1-\gamma_1y\,U_0
\right)
e^{-\gamma_0 y}\,dy
\Bigg].
\end{aligned}
\end{equation}
Thus,
\begin{align}
I_2&=I_{20}+I_{21}a^2+\mathcal{O}(a^4),\\
I_{20}&=K_0+N_0,\\
I_{21}&=K_1+N_1.
\end{align}

The mass accretion rate is
\begin{equation}
\frac{|\dot M|}{mn_R}
=
\frac{4\pi\rho_R^2\alpha(R)}{\sqrt{W}},
\end{equation}
where
\begin{equation}
W\equiv
4z\left(\frac{I_1}{I_2}\right)^2-1
=
W_0+W_1a^2+\mathcal{O}(a^4).
\end{equation}
The coefficients are
\begin{equation}
W_0=
4z_0
\left(\frac{I_{10}}{I_{20}}\right)^2-1,
\end{equation}
and
\begin{equation}
W_1=
4\left[
z_1\left(\frac{I_{10}}{I_{20}}\right)^2
+2z_0
\frac{I_{10}I_{11}I_{20}-I_{10}^2I_{21}}{I_{20}^3}
\right].
\end{equation}
Expanding $W^{-1/2}$ and using
$\rho_R^2\alpha(R)=R_s^2\alpha_0+C'a^2+\mathcal{O}(a^4)$, we obtain
\begin{equation}
\frac{|\dot M|}{mn_R}
=
4\pi
\left[
\frac{R_s^2\alpha_0}{\sqrt{W_0}}
+
\left(
\frac{C'}{\sqrt{W_0}}
-\frac{R_s^2\alpha_0W_1}{2W_0^{3/2}}
\right)a^2
\right]
+\mathcal{O}(a^4).
\label{eq:MdotMaxwellFinal}
\end{equation}
Equation~\eqref{eq:MdotMaxwellFinal} is the $\mathcal{O}(a^2)$ Maxwell--J\"uttner mass accretion rate used in the main text.

\section{Truncation-error curves}
\label{app:error}
\begin{figure}[htbp]
    \centering
    \includegraphics[width=\columnwidth]{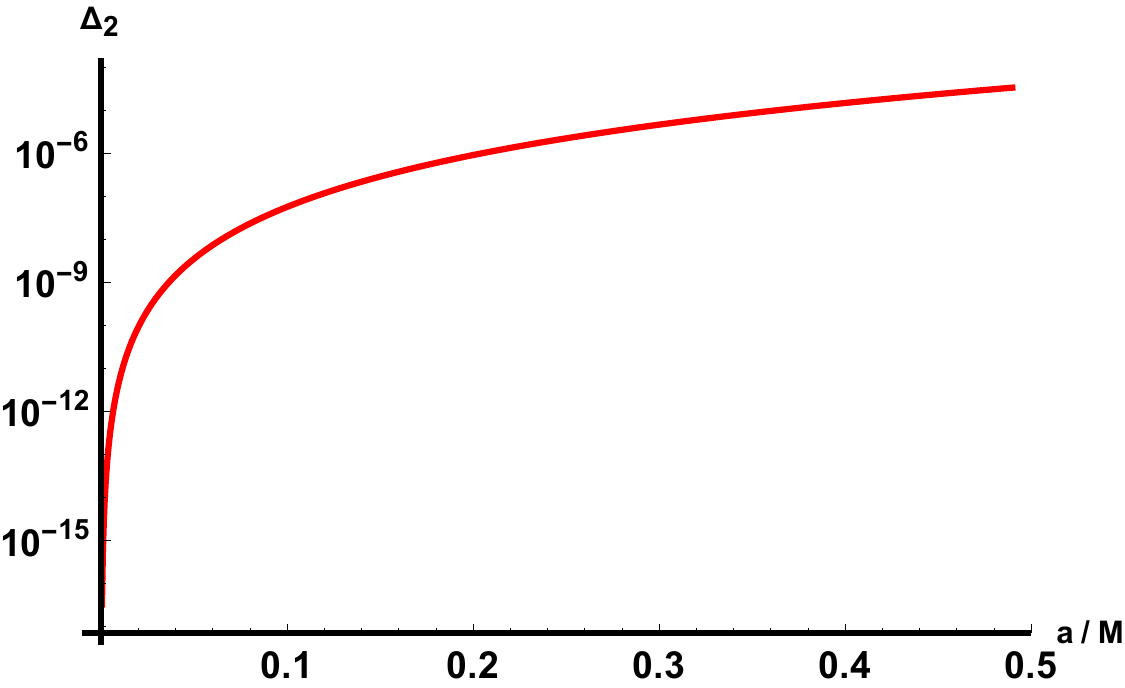}
    \caption{Relative truncation error $\Delta_2$ defined in
    Eq.~\eqref{eq:error}, shown for the monoenergetic model with angular
    momentum. The vertical axis is logarithmic. The error increases
    monotonically with $a$ and reaches $3.33\times10^{-5}$ 
    at the observational upper bound $a/M=0.49$.}
    \label{fig:DeltaError}
\end{figure}

\begin{figure}[htbp]
    \centering
    \includegraphics[width=\columnwidth]{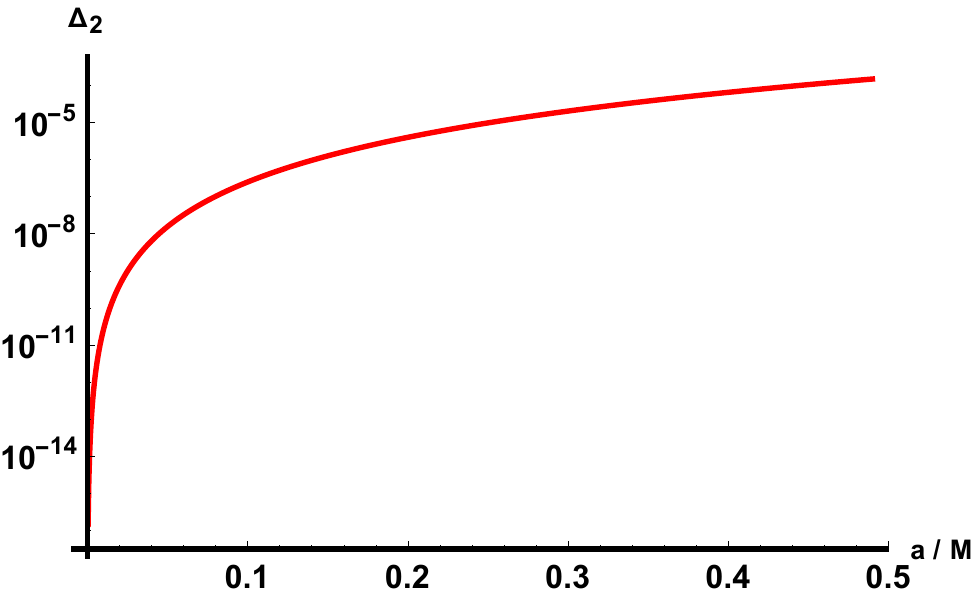}
    \caption{Relative truncation error $\Delta_2$ defined in
    Eq.~(\ref{eq:error}), shown for the Maxwell--J\"uttner model with
    angular momentum. The vertical axis is logarithmic. The error increases
    monotonically with $a$ and reaches $1.49\times10^{-4}$ 
    at the observational upper bound $a/M=0.49$.}
    \label{fig:MaxwellError}
\end{figure}

This appendix collects the relative truncation-error curves for the two
distribution functions.  In both cases the expanded rate is obtained from
the $\mathcal O(a^2)$ expressions, Eq.~(\ref{eq:DeltaMdotExpand2}) and 
(\ref{eq:MaxwellMdotExpand}), evaluated with $R^{(2)}$, while the exact rate is 
computed from Eq.~(\ref{eq:DeltaMdot2}) and (\ref{eq:MaxwellMdot}) at
$R_{\rm exact}$.  The environmental parameters are those adopted in
Sec.~\ref{sec:Vlasov Gas Accretion In The Simpson-Visser Background}.
Figure~\ref{fig:DeltaError} shows $\Delta_2$ for the
monoenergetic model; it increases monotonically with $a$ and reaches
$3.33\times10^{-5}$ at $a/M=0.49$.  Figure~\ref{fig:MaxwellError} shows the
same quantity for the Maxwell--J\"uttner model, with a maximum of
$1.49\times10^{-4}$.  Both vertical axes are logarithmic, and the
residuals are consistent with the expected $\mathcal O(a^4)$ truncation
error.

\nocite{*}

\bibliography{apssamp}

\end{document}